\documentclass[twocolumn,amsmath,amssymb,showpacs,pre,superscriptaddress]{revtex4}
\usepackage{hyperref}
\usepackage[latin1]{inputenc}
\usepackage{vmargin}
\usepackage{graphicx}
\usepackage{rotating}
\usepackage{wrapfig}
\usepackage[dvips]{epsfig}

\newcommand{\eq}{Eq.\,}
\newcommand{\tab}{Tab.\,}
\newcommand{\fig}{Fig.\,}
\newcommand{\figs}{Figs.\,}

\begin{document}
\title{Simulation of Claylike Colloids}
%
%
%
%
\author{Martin Hecht}
\affiliation{ICP, University of Stuttgart, Pfaffenwaldring 27, 70569 Stuttgart, Germany}
\author{Jens Harting}
\affiliation{ICP, University of Stuttgart, Pfaffenwaldring 27, 70569 Stuttgart, Germany}
\author{Thomas Ihle}
\affiliation{ICP, University of Stuttgart, Pfaffenwaldring 27, 70569 Stuttgart, Germany}
\affiliation{North Dakota State University, Department of Physics, Box 5566, Fargo, ND 58105-5566}
\author{Hans J. Herrmann}
\affiliation{ICP, University of Stuttgart, Pfaffenwaldring 27, 70569 Stuttgart, Germany}
%
%
%
%
%
%
\date{\today}

\begin{abstract}
\noindent
\textbf{Abstract.}
We investigate properties of dense suspensions and sediments of
small spherical silt particles by means of a combined Molecular Dynamics
(MD) and Stochastic Rotation Dynamics (SRD) simulation. We include
van der Waals and effective electrostatic interactions between the colloidal particles,
as well as Brownian motion and hydrodynamic interactions which are
calculated in the SRD-part.
We present the simulation technique and first results. We have measured
velocity distributions, diffusion coefficients, sedimentation velocity,
spatial correlation functions and we have explored the phase diagram
depending on the parameters of the potentials and on the volume fraction.
\end{abstract}



\maketitle

\begin{section}{Introduction}
\noindent
We simulate claylike colloids, for which in many cases the attractive Van-der-Waals forces are relevant.
They are often called ``peloids'' (Greek: clay-like). The colloidal particles 
have diameters in the range of some nm up to some $\mu$m. In general, colloid science is
a large field, where many books have been published
\cite{Mahanty76,Lagaly97,Shaw92,Morrison02,Schmitz93,Hunter01}.
The term ``\emph{peloid}''
originally comes from soil mechanics, but particles of this size are also important in many
engineering processes. Our model system of Al$_2$O$_3$-particles of diameter $0.5\,\mu$m
suspended in water is an often used ceramics and plays an important role in technical processes.
In soil mechanics\cite{Richter03} and ceramics science\cite{Oberacker01},
questions on the shear viscosity and compressibility
as well as on porosity of the microscopic structure which is formed by the particles,
arise\cite{Wang99,Lewis00}. In both areas, usually high volume fractions ($\Phi > 20\%$) are of interest.
The mechanical properties of these suspensions
are difficult to understand. Apart from the attractive forces, electrostatic repulsion strongly
determines the properties of the suspension.
Depending on the surface potential, which can be adjusted by the pH-value of the solvent,
one can either observe formation of clusters or the particles are stabilized in suspension and
do sediment only very slowly. Hydrodynamic effects are also important for sedimentation experiments.
Since typical Peclet numbers are of order one in our system, Brownian motion cannot be neglected.

In summary, there are many important factors which have to be included into a model which describes peloids
in a satisfying way. Such a model is needed to gain a deeper understanding of the dynamics of
dense colloidal suspensions. A lot of effort has been invested by 
applying different simulation methods, which have their inherent 
strengths but also some disadvantages.
Simplified Brownian Dynamics (BD), such as in the work of 
H\"utter\,\cite{Huetter00} does not contain long-ranged hydrodynamic 
interactions among particles at all.
The computational cost is low, since hydrodynamics is reduced to a 
simple Stokes force and thus large particle numbers can be handled.
BD with full hydrodynamic interactions utilizes a mobility matrix which
is based on the Oseen- or Rotne-Prager-Yamakawa tensor approximations
which are exact in the limit of zero Reynolds number and zero particle 
volume fraction\cite{Petera99, Ahlrichs01}.\\
This technique faces the main problem that the computational effort 
scales with the cube of the particle number due to the inversion of 
matrices.
\\
The lattice Boltzmann (LB) method on the other hand is numerically 
efficient and intrinsically contains hydrodynamic interactions. Ladd and 
Verberg give an overview over the LB method and describe how to include 
stress fluctuations \cite{Ladd01}. Adhikari et al. add noise to their 
model by introducing a noise term for every lattice velocity and node 
\cite{Cates05}. However, the discussion about the correct inclusion of 
thermal fluctuations is still ongoing \cite{Cates05, Ladd05}.
Pair-Drag simulations have been proposed by Silbert et al.\cite{Silbert97a},
which include hydrodynamic interactions in an approximative way. They have focused on suspensions with
high densities (up to $50\,\%$) of uncharged spherical colloidal particles.
Here we use Stochastic Rotation Dynamics (SRD)\cite{Malev99, Malev00}, a recently developed method 
to simulate fluid flow, and combine this with a Molecular Dynamics (MD) simulation
for the suspended particles. SRD is a particle-based method which does 
not show any numerical instabilities, contains thermal fluctuations 
intrinsically and is simple to implement. Many important issues in 
fluctuation fluid dynamics such as sedimentation \cite{Padding04}, 
vesicles in flow \cite{Gompper04c}, polymers in flow \cite{yeomans-2004-ali}, 
reacting fluids \cite{Kapral04} have been addressed very recently using 
this method.
In this paper, first we discuss the main points of the MD-part of our simulation code, second
we present the SRD method in the context of our work, then we describe two alternative
ways of coupling the two parts of the simulation and point out the advantages 
and disadvantages of these two possibilities. After that, we analyze the time
scales which are relevant for the peloids, we want to simulate. Based on the 
insights of this section we show in the following section how to determine 
the simulation parameters. Then we describe how we have tested our simulation 
code and present first results in the following section. Finally in the last 
section we draw a conclusion and summarize shortly the model we have presented. 
\end{section}

\begin{section}{Molecular Dynamics}
\noindent
The colloidal particles in our simulation are represented by three dimensional spheres.
In order to correctly model the statics and dynamics when approaching
stationary states, realistic potentials are needed.
The interaction between the particles is described by DLVO theory\cite{Huetter00,Russel95,Lewis00}.
If the colloidal particles are
suspended in a solvent, typically water, ions move into solution, whereas their counter ions
remain in the particle due to a different resolvability. Thus, the colloidal particle carries
a charge. The ions in solution are attracted by the charge on the particles and form the
electric double layer. It has been shown\,(see \cite{Russel95}), that the resulting electrostatic
interaction between two of these particles can be described by an exponentially screened Coulomb potential
\begin{equation}
 \begin{array}{rcl}
  V_{\mathrm{Coul}} &=& \pi \epsilon_r \epsilon_0
  \left[ \frac{4 k_{\mathrm{B}} T}{z e}
         \tanh\left( \frac{z e \Psi_0}{4 k_{\mathrm{B}} T} \right)
  \right]^2 \\
  && \times \frac{d^2}{r} \exp( - \kappa [r - d]),
 \end{array}
\end{equation}
where $d$ denotes the particle diameter and $r$ is the distance of the particle centers.
$z$ is the charge of the ions, $e$ the elementary charge, $T$ the temperature, $\Psi_0$
denotes the effective surface potential, and $\kappa$ is the inverse Debye screening length.
In addition the behavior is determined by the attractive van der Waals interaction which can
analytically be integrated over the two spheres. This leads to the second part of the
DLVO potential:
\begin{equation}
 \begin{array}{rl}
  V_{\mathrm{VdW}} = - \frac{A_{\mathrm{H}}}{12} &
     \left[ \frac{d^2}{r^2 - d^2} + \frac{d^2}{r^2} \right. \\
            & \;\left. + 2 \ln\left(\frac{r^2 - d^2}{r^2}\right) \right].
  \end{array}
\end{equation}
$A_{\mathrm{H}}$ is the Hamaker constant which involves the polarizability of the particles
and of the solvent. The DLVO potentials are plotted in \fig\ref{fig_Potentials} for six
typical examples with different depth of the secondary minimum. The primary minimum has to
be modeled separately, as discussed below.
\begin{figure}
\epsfig{file=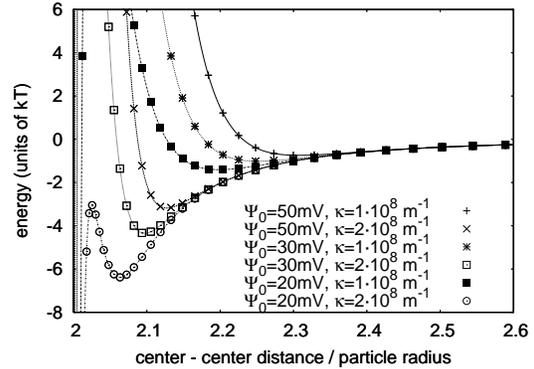,width=\linewidth}
\caption{DLVO Potentials for Al$_2$O$_3$ spheres of $R=0.5\,\mu$m diameter suspended in water.
These are typical potentials used for our simulations as described below. The primary minimum
at $d/R = 2.0$ is not reproduced correctly by the DLVO theory. It has to be modeled separately.
In most of our cases the existence of the secondary minimum determines the properties of the simulated
system.}
\label{fig_Potentials}
\end{figure}
\\
To avoid that the particles penetrate each other, one needs a repulsive force depending
on their overlap. We are using a Hertz force described by the potential
\begin{equation}
  V_{\mathrm{Hertz}} = K (d-r)^{5/2}  \quad  \mathrm{if}  \quad r<d,
\end{equation}
where $K$ could be expressed by the elastic modulus of Al$_2$O$_3$. This would determine
the simulation time step, but to keep computational effort relatively small, we determine
the time step using the DLVO-potentials as described later on and then choose a value for
$K$. Two aspects have to be considered: $K$ has to be big enough so that the particles
do not penetrate each other by more than approximately $10\%$ and it may not be too big, so that
numerical errors are kept small, which is the case when the collision time is resolved
with about 20 time steps. Otherwise total energy and momentum are not conserved very well
in the collision.
\\
Since DLVO theory contains the assumption of linear polarizability, it holds only for
large distances, i.e. the singularity when the two spheres touch, does not exist in reality.
Nevertheless, there \emph{is} an energy minimum about $30\,k_{\mathrm{B}}T$ deep, so that
particles which come that close would very rarely become free again. To obtain numerical
stability of our simulation, we model this minimum by a parabolic potential, some
$k_{\mathrm{B}}T$ deep (e.g. $6\,k_{\mathrm{B}}T$).
The depth of the minimum in our model is much less than in reality, but the probability
for particles to be trapped in the minimum has to be kept low enough so that only few
of them might escape during simulation time.
\\
Long range hydrodynamic interaction is taken into account in a separate simulation for
the fluid as described below. This can only reproduce interactions correctly down
to a certain level. On shorter distances, a lubrication force has to be
introduced explicitly in the molecular dynamics simulation as described in \cite{Schwarzer02}.
The most dominant mode, the so-called squeezing mode, is an additional force
\begin{eqnarray}
  \label{eq_FLub}
  \mathbf{F}_{\mathrm{lub}} &=& -(\mathbf{v}_{\mathrm{rel}},\mathbf{\hat{r}})\mathbf{\hat{r}}
    \frac{6 \pi \eta r_{\mathrm{red}}^2}{r - r_1 -r_2}, \\
  \mathrm{with\quad}r_{\mathrm{red}} &=& \frac{r_1 r_2}{r_1+r_2}
\end{eqnarray}
between two spheres with radii $r_1$, $r_2$ and the relative velocity $\mathbf{v}_{\mathrm{rel}}$.
$\eta$ is the dynamic viscosity of the fluid.
$\mathbf{F}_{\mathrm{lub}}$ diverges if particles touch each other.
Therefore, we limit the force by introducing a minimum radius,
where the force reaches its largest allowed value. The potential is shifted accordingly to smaller
particle distances, so that the maximum force is reached for particles touching each other.
\\
The Hertz force also contains a damping term in normal direction,
\begin{equation}
  \mathbf{F}_{\mathrm{Damp}} =  -(\mathbf{v}_{\mathrm{rel}},\mathbf{\hat{r}})\mathbf{\hat{r}}
    \beta  \sqrt{r - r_1 - r_2},
\end{equation}
with a damping constant $\beta$
and for the transverse direction a viscous friction proportional to the relative velocity of
the particle surfaces is applied.

For the integration of the translational motion we utilize a velocity Verlet algorithm\cite{Allen87}
chap. 3.2.1 to update the velocity and position of particle $i$ according to the equations
\begin{eqnarray}
 {\bf x}_i(t+\delta t) &=& {\bf x}_i (t) + \delta t \,{\bf v}_i(t)+\delta t^2\,\frac{F_i(t)}{m}, \\
 {\bf v}_i(t+\delta t) &=& {\bf v}_i (t) + \delta t \,\frac{F_i(t) + F_i(t+\delta t)}{2 m} .
\end{eqnarray}
For the rotation, a simple Euler algorithm is applied:
\begin{eqnarray}
  {\bf\omega}_i(t+\delta t)&=&{\bf\omega}_i(t) + \delta t \,{\bf T}_i, \\
  {\bf\theta}_i(t+\delta t)&=&{\bf\theta}_i(t) + F({\bf\theta}_i, {\bf\omega}_i, \delta t),
\end{eqnarray}
where ${\bf\omega}_i(t)$ is the angular velocity of particle $i$ at time $t$, ${\bf T}_i$ is
the torque exerted by non central forces on the particle $i$, ${\bf\theta}_i(t)$ is the orientation
of particle $i$ at time $t$, expressed by a quaternion, and $F({\bf\theta}_i, {\bf\omega}_i, \delta t)$
gives the evolution of ${\bf\theta}_i$ of particle $i$ rotating with the angular velocity ${\bf\omega}_i(t)$
at time $t$.
\\
The concept of quaternions\cite{Allen87} is often used to calculate rotational motions in simulations,
because the Euler angles and rotation matrices can easily be derived from quaternions. Using Euler angles
to describe the orientation would give rise to singularities for the two orientations
with $\theta = \pm 90^{\circ}$. The numerical problems related to this fact and the relatively high
computational effort of a matrix inversion can be avoided using quaternions.

We have switched off dissipative forces and checked if the total energy and
each component of the total momentum are conserved. We have verified this for the molecular
dynamics simulation for the simulation of the fluid, and for the coupled simulation separately.
\\
We also checked that our implementation of the molecular dynamics code is correct by
simulating eight large particles with Hertz-repulsion and Coulomb friction in a closed box
at a volume fraction of $\Phi \approx 20 \%$. We checked that the collisions are
realistic, i.e. that the individual angular velocities for two particles interacting in a
non-central collision before and after they have touched are consistent.
\end{section}

\begin{section}{Stochastic Rotation Dynamics (SRD): Simulation of the Fluid}
\noindent
The Stochastic Rotation Dynamics method (SRD) introduced by Malevanets and Kapral \cite{Malev99, Malev00}
 is a promising tool for a coarse-grained description of
a fluctuating solvent, in particular for colloidal and polymer suspensions. The method is also
known as ``Real-coded Lattice Gas'' \cite{Inoue02} or as ``multi-particle-collision dynamics'' (MPCD)
\cite{Gompper04}.
It can be seen as a ``hydrodynamic heat bath'', whose details are not fully resolved but which
provides the correct hydrodynamic interaction among embedded particles\cite{Lamura01}.
SRD is especially well suited for flow problems with Peclet numbers of 
order one and Reynolds numbers on the particle scale between 0.05 and 20 
for ensembles of many particles\footnote{For low Peclet numbers Brownian motion dominates and sedimentation takes place very slowly.
The simulations require a huge number of time steps.
Then Brownian simulation (BS), including short range hydrodynamics interactions, might be a more suitable
tool, since not the complete velocity field has to be calculated. For very high Peclet numbers,
SRD becomes inefficient due to the necessary averaging. For high Reynolds numbers a
small time step and high spacial resolution would be necessary, which increases the computational
effort extremely.}.
The numerical effort scales linearly with the number of embedded 
colloidal particles unlike in Brownian Dynamics, and only one random 
number per node (for the choice of the rotation matrix) is needed in 
contrast to fluctuating lattice-Boltzmann. 
For Peclet-numbers of order one, about three to five SRD-particles 
are required per box (or node) whose positions and velocities can be 
seen as the degrees of freedom in that node. In three dimensions this 
amounts to 18 to 25 variables per node which is similar to the 15 or 19 
speed lattice-Boltzmann method. \\
While the LB method might be slower than SRD in the regime of large 
thermal fluctuations it has the advantage that it can be used for almost 
arbitrarily high Peclet-numbers just by  reducing the amplitude of the 
noise. To reduce the noise in SRD, a huge number of fluid-particles per 
node has to be used which makes the method inapplicable at 
Peclet-numbers higher than about 20. Fortunately it has been shown by 
Louis and Padding\cite{Padding04}, that basic properties of sedimentation such as the 
main settling speed are hardly affected by thermal noise.
 
The method is based on so-called fluid particles with continuous positions and velocities which
follow a simple, artificial dynamics.

The system is coarse-grained into cubic cells of a regular lattice with
no restriction on the number of particles in a cell. The evolution of
the system consists of two steps: streaming and collision. In the
streaming step, the coordinate of each particle is incremented by its
displacement during the time step. Collisions are
modeled by a simultaneous stochastic rotation of the relative velocities
of every particle in each cell. The dynamics
is explicitly constructed to conserve mass, momentum, and energy, and the
collision process is the simplest consistent with these conservation laws.
It has been shown that there is an $H-$theorem for the dynamics
and that this procedure yields the correct hydrodynamic equations
for an ideal gas\cite{Malev99}.

Consider a set
of $N$ point-particles with (continuous) coordinates ${\bf r}_i(t)$ and
velocities ${\bf v}_i(t)$.
In the streaming step, all particles are
propagated simultaneously by a distance ${\bf v}_i\tau$, where $\tau$ is the
value of the discretized time step. For the collision step, particles are
sorted into cells, and they interact only with members of their own cell.
Typically, the simplest cell construction consisting of a hyper cubic grid
with mesh size $a$, is used.
The collision step consists of an independent random rotation
of the relative velocities ${\bf v}_i-{\bf u}$, of the particles in each cell,
where the macroscopic velocity ${\bf u}({\bf \xi},t)$ is the mean velocity of the
particles in the cell with coordinate ${\bf \xi}$.
The local temperature $T({\bf \xi},t)$ is defined via the mean square deviation
of the particle velocities from the mean velocity in a cell.
All particles in a cell are subject to the same rotation, but the
rotation angles of different cells are statistically independent.
There is a great deal of freedom in how the rotation step is implemented
\cite{Malev99,Ihle01},
since, by construction, the local momentum
and kinetic energy are invariant. The dynamics is therefore summarized by
\begin{eqnarray}
\label{eq_move}
{\bf r}_i(t+\tau)&=&{\bf r}_i(t)+\tau\;{\bf v}_i(t), \\
\label{eq_rotate}
{\bf v}_i(t+\tau)&=&{\bf u}[{\bf \xi}_i(t+\tau)]+{\bf \omega}[{\bf \xi}_i(t+\tau)]\\
\nonumber && \cdot\{{\bf v}_i(t)-{\bf u}[{\bf \xi}_i(t+\tau)]\},
\end{eqnarray}
where ${\bf \omega}({\bf \xi}_i)$ denotes a stochastic rotation matrix, and ${\bf \xi}_i$
is the coordinate of the cell occupied by particle $i$ at the time of the
collision. ${\bf u}({\bf \xi})\equiv \frac{1}{M}\sum\limits_{k\in\xi}{\bf v}_k$ is
the mean velocity of the particles in cell ${\bf \xi}$. ${\bf \omega}$ is taken to
be a rotation by an angle $\pm\alpha$, with probability $1/2$. We are using rotations
about the three coordinate axes with $\alpha=\pm 90^{\circ}$, because these are the
most simple rotation matrices one can imagine in 3D, since they only contain
entries taken out of $\{0,\,\pm1\}$. This has been suggested by M. Strau{\ss} in \cite{Ihle03c}.
In every time step for each cell one of these 6 possibilities is chosen with equal probability $1/6$.
However, any stochastic rotation matrix consistent with detailed balance can be used.

In order to remove low temperature anomalies and to achieve exact Galilean-invariance, we use
a modification of the original algorithm \cite{Ihle01}:
all particles are shifted by the {\it same}
random vector with components in the interval $[-a/2,a/2]$ before the collision step.
Particles are then shifted back by the same amount after the collision.
The random vectors of consecutive iterations are uncorrelated.
Ihle and Kroll have discussed in Ref.\cite{Ihle03a, Ihle03b} why this simple procedure works and shown
that it leads to transport coefficients independent of an imposed homogeneous flow field.
In Ref. \cite{Ihle04} and \cite{Yeomans03} analytical calculations of the transport
coefficient in this method are presented.

Two different methods to couple the SRD and the MD simulation have been introduced in the literature.
We have implemented them both and we are using them depending on what we plan to measure.
The first one\cite{Inoue02} is much more accurate in resolving the local velocity field around the
colloidal particles. Lubrication effects are reproduced well by this coupling method. The second
one\cite{Falck04} resolves the velocity field only down to a length scale of the particle diameter.
On the other hand the method becomes much faster because of the lower resolution.
In both coupling methods the long range hydrodynamic interactions are reproduced.
\end{section}

\begin{section}{Coupling I: Placing Fluid Particles Outside of Colloidal Particles}
\noindent
In the combined MD and SRD simulation the fluid particles have to interact with the colloidal particles
and transfer momentum from one to the other part of the simulation. One possibility to do this is, as
suggested by Inoue et al.\cite{Inoue02} to
check after each streaming step of a fluid particle $i$, if its new position
$x_i(t+\tau)$ is within a colloidal particle and if yes, to modify its position and velocity.
In this coupling step total momentum has to be conserved, which means, that when modeling the
``collision'' between the fluid particle and the colloidal particles, one has to make sure that the
change of momentum of the fluid particle is transfered to the suspended particle.
The calculations described in the following are done in a frame fixed on the colloid
particle. 

One can think of several different methods to assign a new position to the fluid particle,
which have been shown to work properly:\\
1) place it on the shortest distance to the surface of the colloidal particle and move it
with its new velocity half of a time step,\\
2) calculate the point and the exact time when the fluid particle has entered into the
colloidal particle and move it back to there. Then choose a new velocity and move the fluid
particle with the new velocity for the remainder of the time step.\\
Both methods turned out to work, where the second one is more accurate but more computationally
intensive as well. Just to place the fluid particle directly on the surface and move it
again in the next time step turned out to produce an increase of the fluid particle density
around the colloidal particle. Anomalies in the fluid temperature could also be found when
the fluid particles were placed directly on the colloid surface.

To increase stability of the simulation the idea is not to conserve energy in every
single collision, but to use a thermostat and choose the new velocities according to a
given distribution. The new velocities should point from the colloid surface to the
outer area. Since the interior of the colloidal particle usually does not contain any fluid
particles and the velocity distribution next to a colloidal particle should be independent
of neighboring particles, the velocity distribution for the newly chosen
fluid particle velocities has to be the same as if the space inside the suspended particle was filled
with fluid particles. Assume these imaginary fluid particles having the same density and
temperature as in the remainder of the fluid bath. Then, one could evaluate the velocity distribution
for the reflected fluid particles by taking the velocity distribution of the imaginary fluid particles
passing through the colloid surface. But it is a non-trivial task to analytically calculate
this distribution for a spherical area.
However, if the mean free path of the fluid particles is small compared to the diameter
of the colloidal particles, we can safely assume
the colloid surface to be an infinitely extended plane separating the space into
two regions\cite{Inoue02}. Then one finds the following distribution:
\begin{eqnarray}
  p(v_n) &\sim& v_n \exp(-\beta v_n^2), \\
  p(v_t) &\sim& \exp(-\beta v_n^2) \\
  \nonumber && \mathrm{with} \; \beta = \frac{m_f}{2 k_{\mathrm{B}}T},
\end{eqnarray}
where $v_n$ is the normal component and $v_t$ is the tangential component of the fluid particle velocity in
the frame fixed to the surface of the large particle.
$m_f$ is the mass of a fluid particle. In the following sections we describe how $m_f$
has to be chosen. $T$ is the temperature
to which this thermostat is adjusted and the whole system will
adopt this temperature after a transient time. The tangential component can be obtained by
computing $\sqrt{x_1^2+x_2^2}$ of two independent and Gaussian distributed random variables.

Since the fluid particles of the SRD are artificial particles within the context of this
mesoscopic simulation method, their mean free path and their momentum are different from the
corresponding values for single solvent molecules. Because of this, there is a depletion
force acting on colloidal particles which is much larger than in reality. Depletion forces are only
relevant in systems with very big molecules, e.g. polymer solutions with added small particles or
binary mixtures of particles with clearly separated diameters. There, each of the small particles
carries a considerable momentum - which is also the case in the SRD simulations.
Nevertheless, unrealistically high depletion forces can be suppressed by reflecting fluid particles
many times: If after the collision step the fluid particle is placed in another colloidal particle,
the collision step is repeated for that colloidal particle and so on, until the fluid particle reaches a
position outside any colloidal particle or until a maximum number $N_{\mathrm{max}}$ of
collisions has been calculated through. We have measured the depletion force and found out that
a limit of $N_{\mathrm{max}}\sim 10$ is a good compromise between computational speed and accuracy.
The depletion force does not decay substantially stronger if the limit is increased, but the computational effort
still grows with $N_{\mathrm{max}}$ (at most linearly), because some fluid particles are trapped in a
small gap between two colloidal particles and jump from one to the other. This in fact would still decrease
the depletion force, but in the mean time the calculation for the remaining system is interrupted
until finally eventually one single fluid particle is reflected the very last time. It is obvious
that this scenario can easily be truncated. The remaining depletion force can be neglected at
least in the cases where strong attractive van der Waals forces or strong repulsive electrostatic
forces are present.
\end{section}

\begin{section}{Coupling II: Rotating Velocities of the Colloidal Particles}
\noindent
A second possibility to couple the SRD and the MD simulation is to sort the colloidal particles into
the SRD boxes and include their velocity in the rotation step. This technique has been used to model 
protein chains suspended in a liquid\cite{Falck04, Gompper04b}. The mean velocity in
each cell has then to be weighted  with the mass of the particle (because the mass of colloid
particles differs at least by one order of magnitude from the one of the fluid particles and their
inertia dominates the flow field next to it). The calculation of
${\bf u}[{\bf \xi}_i(t+\tau)]$
in \eq(\ref{eq_rotate}) is modified to
\begin{eqnarray}
\label{eq_rotateMD}
{\bf u}({\bf \xi})\equiv \frac{1}{M}\sum\limits_{k\in\xi}{\bf v}_k m_k,
\end{eqnarray}
where we sum over all colloid and fluid particles in the cell. $m_k$ is the mass of the
particle with index $k$ and $M \equiv \sum\limits_{k\in\xi}m_k$ is the total mass contained
in the cell.

The coupling acts on the center of mass of the colloidal particles
and affects only the fluid particles within the same cell. This means, to affect the same area
of the flow field like in reality, one has to choose the cells to be of the same size as the
colloidal particles. Obviously, the mesh size is drastically larger than in the first coupling
method and the flow field cannot be resolved in detail. The fact that colloidal particles push
away the solvent as well as depletion and lubrication forces cannot be reproduced at any level.

\end{section}

\begin{section}{Time Scale Analysis}
\noindent
Our system contains many different, let us say $L$, time scales, which differ by several orders of magnitude
making {\it brute force} numerical simulations very time-consuming or even impossible.
These time scales can be used to define $L-1$ dimensionless characteristic numbers, such as the Reynolds- or
the Peclet number as the ratio of two time scales.
If one can manage to adjust the simulation parameters such that all these characteristic numbers are the same
as in the experiment, the simulations should be able to exactly reproduce the dynamical behavior of the real
system. Of course, therefore one has to change quantities like the temperature or the viscosity of the
fluid.
\\
Often, it is sufficient to reproduce only a few of all characteristic numbers exactly, i.e. only those which
are believed to be significant for the behavior. For example, in sedimentation processes where the Reynolds
number is much smaller than unity, it may be modified to another value, which still fulfills the condition
of being much smaller than one. In both cases the Stokes limit is a valid
approximation. As a general rule of thumb, dimensionless numbers of order one are important to be reproduced
since they represent two competing dynamical effects.
The reason to modify the other ``insignificant'' numbers is to reduce the ratio of the largest to the
smallest time scale which determines the numerical effort.
In order to decide which are the dimensionless numbers that can be safely modified without changing the
physics too much, a detailed analysis of the different time scales is needed.
\\
We start with the largest scales.
After some time an isolated  spherical particle sedimenting in a liquid reaches the so-called Stokes velocity,
\begin{equation}
v_S=\frac{2}{9}\, \frac{R^2 g}{\nu}\,\left(\frac{\rho_m}{\rho_w}-1\right).
\end{equation}
$\nu$ is the kinematic viscosity, $g$ denotes gravity, $\rho_m$ is the mass density of the particle, $\rho_w$ the
mass density of the solvent. This velocity is obtained from the force balance between buoyancy and weight of the
particle,
$F_G=4\pi (\rho_m-\rho_w) g R^3/3$ and the drag-force  in a viscous liquid,
$F_D=6\pi \nu \rho_w R v$.
\\
The drag-force $F_D$ also defines the mobility $\mu=v/F_D=1/(6\pi \nu \rho_w R)$ of a spherical particle.
The time for a particle to move a distance of its diameter, $2R$, is denoted by
\begin{equation}
\tau_S=\frac{2 R}{v_S}=\frac{9 \nu}{R g \left(\frac{\rho_m}{\rho_w}-1\right)}.
\end{equation}
By means of the Einstein-relation $D=\mu k_{\mathrm{B}} T$ we obtain the diffusion constant $D$ for the particle,
\begin{equation}
\label{eq_Einstein}
D= \frac{k_{\mathrm{B}} T}{6 \pi \nu \rho_w R}.
\end{equation}
The mean square displacement of a diffusing particle in each dimension $i$ is given by
\begin{equation}
<x_i^2(t)>=2Dt\,,
\end{equation}
hence, the time the particle needs to diffuse a distance of $2R$ is of the order of
\begin{equation}
\tau_D=\frac{2 R^2}{D}=\frac{12 \pi \nu \rho_w R^3}{k_{\mathrm{B}} T},
\end{equation}
which we call diffusion time.
\\
The ratio $\tau_D/\tau_S$ measures the importance of Brownian motion in the system and is called Peclet number,
$Pe=\frac{\tau_D}{\tau_S}$. It turns out to be close to unity here. Inserting the definitions for
$\tau_D$ and $\tau_S$, one notices that $Pe$ depends on the fourth power of the radius $R$,
\begin{equation}
Pe=\frac{v_S R}{D}
=\frac{F_G R}{k_{\mathrm{B}} T}
=\frac{4\pi g R^4 (\rho_m-\rho_w)}{3 k_{\mathrm{B}} T}.
\end{equation}
Let us consider another time proportional to $\tau_D$: we assume a regular three-dimensional, cubic array of
spheres which are separated by gaps of size $R/2$. Then, the volume concentration of this suspension is in
the intermediate regime, $\phi=0.268$. The time one sphere diffuses the distance of a gap is given by
$\tau_G=\tau_D/16$.
\\
Now, let us discuss another important time called the particle relaxation time, which is related to how long
it takes the particle to react to an imposed force, i.e. this time measures the inertia effects.
Consider Newton's equation for a particle of mass $m$ subject to a force $F$ and a friction coefficient $\xi$,
$m\frac{\partial v}{\partial t}=-\xi v +F$. Expanding the velocity $v$ around the stationary state,
$v=v_S+\delta v$,
gives
\begin{equation}
\frac{\partial \delta v}{\partial t}=-\frac{\xi}{m} \;\delta v,
\end{equation}
which leads to an exponential decay on a time scale $\tau_P=m/\xi$.
Identifying the friction $\xi$ with $1/\mu$ and inserting the mass leads to
\begin{equation}
\label{Eq_Def_Tau_p}
\tau_P=\frac{2}{9}\,\frac{R^2}{\nu}\,\frac{\rho_m}{\rho_w}.
\end{equation}
\\
Now we consider a very short time scale $\tau_F$, the time fluid momentum diffuses a distance $2R$,
i.e. $(2R)^2=2\nu \tau_F$ leading to
\begin{equation}
\tau_F=\frac{2 R^2}{\nu}
\end{equation}
which helps defining the particle Reynolds number as
\begin{equation}
Re=\frac{\tau_F}{\tau_S}=\frac{R v_S}{\nu}.
\end{equation}
\\
Finally, we have to discuss another important short length scale due to a short range potential among the
colloidal particles. This scale usually determines the maximum time step in Molecular Dynamics.
Guided by the analogy to a harmonic oscillator with frequency $\omega=\sqrt{k/m}$, we replace the spring
constant $k$ with the second derivative of the inter-particle potential $\partial^2 V(R)/\partial R^2$ and
use the period of this oscillation to define the interaction time scale
\begin{equation}
\label{eq_Tau_V}
\tau_V=\frac{2 \pi}{\omega}=2\pi\sqrt{\frac{ml^2}{A_H}},
\end{equation}
where we approximate the derivative of the potential by means of the Hamaker-constant $A_H$ as a typical
size of the potential and a typical distance $l$ such as the distance between the surface of the particle
and the primary potential minimum due to the combined effect of van der Waals attraction and screened
Coulomb repulsion.
Comparison of $\tau_V$ and $\tau_P$ can answer the question, whether the oscillations of two particles
around the primary or secondary minimum are visible or whether the creeping or over-damped case
is realized where friction is dominating over inertia.
Analyzing a harmonic oscillator with damping constant $\xi$ one finds that creeping being established at
\begin{equation}
\label{eq_TauVTauP}
\tau_P\leq \frac{\tau_V}{4\pi}.
\end{equation}
In using this relation a lubrication force described in \eq(\ref{eq_FLub}) has to be taken into account.
This force is proportional to the difference of normal velocities of two approaching particles and in this sense
it can be seen as an additional contribution to the friction coefficient.
It becomes huge at short inter-particle distances $d$ and it will turn out later that even without this addition
all particles considered here are well inside the creeping regime due to the large friction in water.
This is the justification that so far many people used Brownian Dynamics (BD)
for this system instead of Molecular Dynamics (MD)\cite{Huetter00, Blaak04}.
In our situation, including thermal fluctuations and full hydrodynamics consistently is easier to do in
Molecular Dynamics. Moreover, with our parameters the MD is at least competitive or even faster
than previous BD calculations.
\end{section}

\begin{section}{Similarity Considerations and Determination of Simulation Parameters}
\begin{subsection}{Introduction}
The determination of parameters for a mesoscopic model to quantitatively compare with experiment 
is a non-trivial task.
Typical values of the parameters in an experiment are listed in \tab\ref{tab_Parameters}.
\begin{table*}
\begin{tabular}{|l|l|}
\hline
Particle radius $R$ & $0.4\, \mu$m \\
Temperature $T$    & $300\,$K \\
mass density of particle $\rho_m$ & $3.9\cdot 10^3\, \mathrm{kg/m}^3$ \\
mass density of water $\rho_w$ & $1.0\cdot 10^3\, \mathrm{kg/m}^3$ \\
Boltzmann constant $k_{\mathrm{B}}$ & $1.38\cdot 10^{-23}\,$J/K \\
kin. viscosity of water $\nu$ & $10^{-6} \,\mathrm{m}^2\mathrm{/s}$ \\
gravity $g$  & $9.81\, \mathrm{m/s}^2$ \\
Hamaker constant $A_H$ of $Al_2O_3$ in $H_2O$ & $4.76\cdot 10^{-20}\,$J \\
distance to primary minimum $l$ & $0.008\,\mu$m \\
\hline
\end{tabular}
\caption{Parameters for the simulation}
\label{tab_Parameters}
\end{table*}
For these values of the parameters in the experiment all the time scales defined in the
previous section are calculated and listed in \tab\ref{tab_Timescales}.
\begin{table*}
\begin{tabular}{|l|l|l|l|l|l|}
\hline
$\tau_S$   & $\tau_D$   & $\tau_G$   & $\tau_V$     & $\tau_F$      & $\tau_P$      \\ \hline
$0.791\,$s & $0.582\,$s & $49.4\,$ms & $7.45\,\mu$s & $0.320\,\mu$s & $0.139\,\mu$s \\ \hline
\end{tabular}
\caption{time scales which arise in a system characterized by the
  parameters listed in \tab\ref{tab_Parameters}}
\label{tab_Timescales}
\end{table*}
This tells us that the Peclet number is $Pe=\tau_D/\tau_S=0.74$, and we have a competition
between convection due to gravity and Brownian motion.
The particle Reynolds number is very small, i.e. $Re=\tau_F/\tau_S=4.0\cdot 10^{-7}$.
The ratio of $\tau_V$ to $\tau_P$ is larger than $4\pi$, hence oscillations of particles
in their short range potentials are over-damped, already without considering lubrication forces.
We get $\tau_P\ll \tau_G$, since the particles are well relaxed before they hit
each other due to Brownian motion. $\tau_F\ll \tau_D$, hence the transport of momentum through
the fluid is much faster than if transported directly by the particle.
These are the dynamical characteristics which have to be preserved by any parameter changes,
in particular, the Peclet-number has to be kept exactly the same.
Of course, the static properties such as the ratio of kinetic energy $\sim k_{\mathrm{B}} T$ to the
potential energies, $\sim mgR$ and $\sim A_H$ have to be kept the same too. 
\\
However, using identical parameters as shown in \tab\ref{tab_Parameters} 
in an MD simulation would require of the order of $10\tau_S/\tau_P\approx 5\cdot 10^7$ iterations to see 
sufficient progress in the sedimentation process. This is an unacceptably high numerical effort, which must be 
reduced without  significantly changing the physics of this process. First, we now show how to choose the 
parameters for a simulation using the coupling method I. After that, we describe what has to be changed    
using the coupling II.
\end{subsection}

\begin{subsection}{Determination of the parameters for coupling I}
We start by choosing reasonable parameters for the hydrodynamic part of the code, i.e.
Stochastic Rotation Dynamics (SRD), since this is time-consuming and the most storage-intensive part of
our simulation. For the moment we
keep the particle radius constant at $R=0.4\,\mu$m. Let $a$ be the lattice constant of the SRD grid.
By choosing $a=R/2$ a spherical particle covers about $34$ boxes which is a sufficient resolution of
the particle. 
We get $a=0.2\,\mu$m.
\\
We use an average number of $M=2.5$ fluid particles per box, which leads to $6M=15$ real numbers
(3 velocity and 3 position coordinates in 3 dimensions) to be stored for every box.
A larger $M$ would reduce Brownian motion and increase CPU-time and storage requirements. Using a smaller number
leads to a very long effective mean free path of the fluid particles (sometimes there is only one particle per box
and no collision takes place), which results in a large viscosity and a bad resolution of the flow field
around the colloidal particles.
\\
Next, we choose the ratio of the mean free path $\lambda=\tau\sqrt{k_{\mathrm{B}} T/m_f}$ to the lattice constant.
$m_f$ is the mass of the fluid particle and $T$ the effective temperature of the fluid particles which
can differ by several orders of magnitude from the real temperature of the experiment as will be explained
later. In Ref. \cite{Ihle01}
it was discovered that a ratio $\lambda/a$ smaller than 0.5 leads to anomalies in the model, which can be
corrected by a random shift of the lattice prior to every rotation.
Here, we set $\lambda=0.6\,a=0.12\,\mu$m to have sufficient resolution of the flow and random shifts
are not needed.
\\
The rotation angle $\alpha$ is taken to be $90^{\circ}$ because this gives the most simple rotation matrix.
The exact expression for the shear viscosity 
for $\alpha=90^{\circ}$ is given by \cite{Ihle04}
\begin{equation}
\label{NUF1}
\nu=\frac{a^2}{18 \delta t}\left(1-\frac{1-{\rm e}^{-M}}{M} \right)
+\frac{k_{\mathrm{B}} T\,\delta t}{4 m_f}\,\frac{M+2}{M-1}.
\end{equation}
Inserting $M=2.5$ and expressing temperature by means of $\lambda$ it follows
for our choice of parameters that
\begin{equation}
\label{NUF2}
\nu=0.3052 \frac{a^2}{\delta t}.
\end{equation}
In order to reproduce the same diffusion coefficient as seen in experiments, $\delta t$ has to be determined
by means of the Einstein relation,
\begin{equation}
D=k_{\mathrm{B}} T\,\mu = \frac{k_{\mathrm{B}} T}{6\pi \nu \rho_w R}.
\end{equation}
Setting $\rho_w=M\,m_f/a^3$, using $\nu$ from \eq(\ref{NUF2})
and expressing $k_{\mathrm{B}} T/m_f$ by means of $\lambda$
one finds
$\delta t=0.025 a^3/(DR)$.
Inserting the diffusion coefficient expected in reality from the  Einstein relation,
$D=5.49\cdot 10^{-13}\,\mathrm{m^2}/\mathrm{s}$ we arrive at a time step $\delta t=0.91\,$ms for the
SRD algorithm.
This time step is of course too large to resolve the motion of colloidal particles due to inter-particle forces
and friction. Hence, a two-step method is needed: The trajectory for the colloidal particles is integrated by 
another, smaller time step $\delta t_M$. This also means, that the extensive SRD-procedure is only applied every
$\delta t/\delta t_M$th iteration of the MD-algorithm, thus reducing the required computer power substantially.
\\
The way parameters are derived implicitly means that we keep $\tau_S$ and $\tau_D$ as in reality.
This corresponds to $\tau_S/0.91\,\mathrm{ms}=869 $ SRD-iterations until a colloidal particle has fallen
down by one diameter $2R$ which is affordable.
The kinematic viscosity in the simulation is much smaller than in nature
($\nu_{\mathrm{model}}=1.34\cdot 10^{-11} m^2/s$).
\\
Next, one has to check what happens to the particle relaxation time $\tau_P$.
The requirement is, that it should be much larger than the one given in \tab\ref{tab_Parameters}
(in order to increase numerical efficiency) and on the other hand it should still be smaller
than $\tau_G$ to ensure that particles can relax between consecutive collisions caused by thermal motion.
Following \eq(\ref{Eq_Def_Tau_p}), we obtain $\tau_P=7.69\,$ms.
This is an acceptable value: it is much larger than the $0.139\,\mu$s seen in reality and still
smaller than $\tau_G=49.4\,$ms. Therefore it needs 7 SRD-steps to relax a particle which
means that the process still can be resolved.
\\
Considering momentum transport in the fluid versus direct transport:
During time $\tau_D$, momentum in the fluid is transported a distance $x^2=2\nu \tau_G=33.10 a^2$,
i.e. $x=5.75 a=2.88 R$. Hence, momentum transport in the fluid is only slightly faster than by diffusive
transport, which is still acceptable, even though in the real system it is much faster.
This is reflected in a Reynolds number which is larger by a factor of
$10^{-6}/\nu_{\rm model}=0.746\cdot 10^{5}$ in the simulation, i.e. $Re=2.9\cdot 10^{-2}$.
This again reflects the fact that the SRD-model is efficient only if Peclet- and Reynolds number
are in the range between $0.05$ and $20$.
\\
Now, the gravity constant $g$ of the model has to be determined requiring that the Stokes velocity is
the same as given in \tab\ref{tab_Parameters}.
Since thermal convection of the fluid is not important for our simulation, we can neglect gravity on the
fluid particles. Therefore, there is no buoyancy force in the simulation.  We can correct for that
by assuming a smaller gravity constant modified by the density ratio of colloid material and fluid. We find
\begin{eqnarray}
\nonumber
g_{\rm model} &=& g_{\rm real} \frac{\nu_{\rm model}}{\nu_{\rm real}}
  \left(1-\frac{\rho_w}{\rho_m}\right)\\
  &=& 9.78\cdot 10^{-5} \frac{\mathrm{m}}{\mathrm{s}^2}
\end{eqnarray}
As mentioned above, not only the viscosity, but also the temperature in our simulation may be different
from the one in nature. To see that we calculate the ratio $\Lambda=\rho_m/k_{\mathrm{B}} T$.
In nature we have $\Lambda=0.942\cdot 10^{24} \,s^2/m^5$.
In the model we get $\Lambda_{\rm model}=3.9\,M m_f/(a^3 k_{\mathrm{B}} T)$
where we express $k_{\mathrm{B}} T / m_f$
by means of the mean free path and the time step $\lambda^2/(\delta t)^2$.
One finds that $\Lambda_{\rm model}$ is scaled by a factor of $7.44\cdot 10^4$.
The static features have to be reproduced by the model, and therefore we have to
keep the ratio of kinetic and potential energy $k_{\mathrm{B}} T/A_H$ constant.
This means the ratio $\rho_m/V_{\mathrm{Pot}}$ and especially $\rho_m/A_H$ has
also to be scaled by this factor.
We use $A_H=4.76\cdot 10^{-20}\,\mathrm{J}/(7.44\cdot 10^4)$ in the model, corresponding to new
$A_H = 8.61\cdot 10^{-25}\,\mathrm{J}$. From
\eq(\ref{eq_Tau_V}) we get a
scaled $\tau_V$ of $2.03\,$ms corresponding to
$\tau_V/\tau_P=0.264$ (which is smaller than $4\pi$, see \eq\ref{eq_TauVTauP}).
The unscaled value is 53.6.
The creeping case is restored by the lubrication force, which we have included in the MD simulation,
and which grows for smaller gaps between the particles.
The lubrication force determines the small iteration time step $\delta t_M$ for the MD simulation.
We chose $\delta t_M=2\mu\mathrm{s}$, which is about 200 times larger than it would be if all the
original parameters would have been kept and $\min(\tau_V,\tau_P)$, being much smaller, would
determine the time step.
\\
Comparing to the SRD time step we see that every 455 small steps one SRD step is performed.
We need 869 SRD steps and $4.3\cdot 10^6$ MD steps to see a colloidal particle sinking down by one diameter.
The time scales in the simulation are summarized again in \tab\ref{tab_TimescalesCouplingI}.
\begin{table*}
\begin{tabular}{|l|l|l|l|l|l|}
\hline
$\tau_S$   & $\tau_D$   & $\tau_G$   & $\tau_V$     & $\tau_F$      & $\tau_P$   \\ \hline
$0.791\,$s & $0.582\,$s & $49.4\,$ms & $2.03\,$ms   & $22.9\,$ms    & $7.69\,$ms \\ \hline
\end{tabular}
\caption{Time scales in the simulation using coupling method I.}
\label{tab_TimescalesCouplingI}
\end{table*}
\end{subsection}

\begin{subsection}{Determination of the parameters for coupling II}
To simulate the same system with coupling method II, we use the same particle radius
$R=0.4\,\mu\,m$. The lattice constant has now to be chosen differently because the colloidal particles are
coupled to the SRD-simulation as mass points. They have influence on the fluid which is in the same cell,
and therefore the size of the cell can be understood as the volume within which the SRD-simulation
``feels'' the colloidal particles and we choose the lattice constant in a way that the volume of the cell
is equal to the volume of a colloidal particle: $a = 6.25\cdot 10^{-7}\,\mathrm{m}$. A smaller lattice constant
in this context would model smaller colloidal particles in the SRD-part of the simulation. The velocity field
would be resolved better, but since coupling method II does not allow a resolution smaller than
the colloidal particles, one can not expect to gain any information from the fluid simulation on smaller length
scales than the colloidal particle size. Any attempt to increase the resolution of the SRD simulation
would only cause a larger computational effort.
\\
Since we do not modify the Peclet number, we have to choose approximately the same number of fluid
particles per colloidal particle. Since the box size has increased with respect to the coupling method I, we have
to assume more particles per box now. We choose $M = 60$ (which would correspond to two particles per
box in the coupling method I, but since the boxes are much larger now, we can slightly reduce the ratio of
fluid particles per colloidal particle).
\\
We choose $\lambda/a = 0.5$ and use random grid shifts here to avoid that fluid particles interact
too often with the same partners which causes artefacts in their correlation. The rotation angle $\alpha$
is again $90^{\circ}$ to achieve very simple matrices. Following the same procedure
as for coupling method I (\eq\ref{NUF1}) we find a time step for the SRD of
\begin{equation}
  \delta t = 2.05\,\mathrm{ms.}
\end{equation}
According to \eq\ref{NUF2} the viscosity in the simulation results to
$\nu_{\mathrm{model}}=2.29\cdot 10^{-11} \mathrm{m^2/s}$. The gravity constant
therefore has to be rescaled by a factor of $58813$, and the temperature and the potentials
have to be scaled by $43733$\footnote{Where we take care of the fact that we do not apply
gravity to the fluid particles.}.

The resulting characteristic times are shown in \tab\ref{tab_TimescalesCouplingII}.
\begin{table*}
\begin{tabular}{|l|l|l|l|l|l|}
\hline
$\tau_S$   & $\tau_D$   & $\tau_G$   & $\tau_V$     & $\tau_F$      & $\tau_P$   \\ \hline
$0.791\,$s & $0.582\,$s & $49.4\,$ms & $1.56\,$ms   & $14.0\,$ms    & $18.2\,$ms \\ \hline
\end{tabular}
\caption{Time scales in the simulation using coupling method II.}
\label{tab_TimescalesCouplingII}
\end{table*}
$\tau_S$ and $\tau_D$ are again kept as in reality. Now we need $\tau_S/0.00106\,\mathrm{s}=385 $
SRD-iterations until a colloidal particle has fallen down by one diameter, which is much faster than by
using coupling method I.
$\tau_P$ is still smaller than $\tau_G$, but these two times are now of the same order of magnitude.
This reflects the fact that with coupling method II the local flow around the particles cannot be
resolved. Relaxation of the particles between their collisions due to Brownian motion is in this case
less important because lubrication effects can not be seen here. $\tau_F$ is of comparable size, too,
which means that the diffusion of the momentum is now on the same time scale as the particle motion.
This is understandable since the length scales of the colloidal particles and of the fluid boxes have
to be the same.

Momentum is transported 1.2 times faster in the fluid as by the particles themselves. Short range
hydrodynamic interactions which cannot be resolved are in this sense comparable to particle - particle
collisions whereas for long range interactions the slightly faster transport of momentum can reproduce
coarse grained hydrodynamic effects. Again, to model these effects comparable to reality, the
Reynolds number has to be much smaller than unity. We find $Re = 1.77 \cdot 10^{-2}$.

If we include lubrication forces in the MD simulation in order to reproduce at least to some extend
short range hydrodynamics, we have to choose the same MD time step as for coupling method I but
we need approximately fifty per cent less CPU time for the hydrodynamics. Even though
it seems to be a too simplified approach, we can reproduce a volume fraction dependent sedimentation
velocity as will be described in the results section.
\end{subsection}
\end{section}

\begin{section}{Simulation Setup}
\begin{subsection}{Boundary conditions}
\noindent
Most simulations have been performed using periodic boundary conditions 
in all three directions. Then the total
momentum may not change in any simulation step if no external
forces (like gravity) are applied.  If gravity on the colloidal particles is
applied in a system with periodic boundary conditions, this would accelerate the
whole system, since the total force on the center of mass is not vanishing.
In a real system there is friction at the walls and even more important, there
is an equilibrium between hydrostatic pressure acting on the surface of a
given volume and the gravity acting as a body force. Since we
simulate a volume in the center of the suspension we either have to
apply the pressure on the walls or, which is easier, make sure that
in sum the forces on the center of mass of the whole simulated system
vanishes. Therefore, we follow the center of mass, i.e. on particles with
higher density their gravity minus buoyancy has to be applied, so that they move
downward whereas the same force in opposite direction has to be applied to the fluid,
which makes it move upward like in a sedimentation vessel with a closed bottom.

For the following discussion we define that the direction in which eventually
gravity is applied is called $-z$-direction, if a shear force is applied acts in
the $x$-direction.
Using closed boundaries
wall effects may be introduced, e.g. crystallization starts earlier than in the bulk.
This effect could be observed especially when gravity was switched off and only closed boundary
conditions were applied. This is a finite size effect, which is not that strong, if periodic
boundaries are applied. But in the case of gravity being applied, the whole system accelerates.
To face this problem, three possibilities were tested: \\
1) fix the boundaries only in $z$-direction,\\
2) fix the boundaries in $x$ and $y$-direction and apply no-slip for the fluid,\\
3) choose periodic boundaries in all directions and compensate the gravitation on the colloidal
particles with a force in the opposite direction applied on the fluid.
\\
Possibilities 1) and 2) simulate a system close to a wall, in case 1) it is the bottom of a
vessel whereas in case 2) the experiment would be done in a capillary. Possibility 3) turned
out to be the most realistic simulation although it can start to drift, if the compensating
force is not adjusted accurately. Slowly accumulated drifts of the center of mass can be
removed every hundreds of SRD-time steps if necessary.
\end{subsection}

\begin{subsection}{Temperature and thermostat}
We have measured the temperature of the colloidal particles for different setups 
If damping constants are chosen appropriately, the resulting
temperature fits very well the temperature, which we have adjusted for the fluid
by the initial conditions. If we additionally switch on a thermostat which we describe
in the following the measured temperature exactly agrees with temperature adjusted
by the thermostat. When gravity is applied to the
system, particles are accelerated and if in addition periodic boundaries are used, a
thermostat is absolutely needed to remove the extra energy, introduced by the periodic
boundary in $z$ direction in combination with gravity.

Therefore we use a modified version of the thermostat described
in \cite{Allen87} [Chap.~7.4.1, ``Stochastic methods'' p.227 f.].
The thermostat, originally suggested by Heyes\cite{Heyes83},
chooses a random scaling factor $\zeta$ for the velocities from
an interval $[1-\gamma,1+\gamma]$. The scaling of the velocity
is then accepted or rejected according to a Monte Carlo
scheme. 
However, the detailed balance is not fulfilled for the choice of $\zeta$
described in\cite{Allen87}.
In our implementation of the thermostat, we randomly choose an $\epsilon$ in the
interval [$0,\,\gamma$] and apply for $\zeta$ one of the values $1+\epsilon$
 or $\frac{1}{1+\epsilon}$, each of them
with the probability of $\frac{1}{2}$. With one of these values
the velocity is scaled by the Monte Carlo acceptance rate.
Also the temperature in our case is defined slightly different
from\cite{Allen87}:
the mean velocity ${\bf u}$ within one SRD-cell defines the
velocity field of the fluid and gives the hydrodynamic interaction
between the colloidal particles. Therefore it may not be modified by the
thermostat. We only scale the velocity component relative to the
mean velocity: ${\bf v}_i^{new} = \zeta({\bf v}_i - {\bf u}) + {\bf u}$.
The Monte Carlo acceptance rate in our case is given by
\begin{equation}
  \begin{array}{l}
  \zeta^{(3(M-1))} \exp(-(M-1) (\zeta^2-1) T/T^{*}) \\
  \mathrm{with} \quad
  T = \frac{m_f}{2(M-1)k_{\mathrm{B}}} \sum\limits_{i=1}^M ({\bf v}_i - {\bf u})^2,
  \end{array}
\end{equation}
which is the local temperature in the SRD-cell and $T^{*}$ denotes
the temperature to which the thermostat will drive the system. $M$ is
the number of particles in the cell. Note that one has to divide
the total thermal energy in the SRD-cell by $M-1$ instead of $M$ to
calculate the local temperature. This reflects the fact that
the mean velocity ${\bf u}$ in the cell already contains three
degrees of freedom which the particles in the SRD-cell have.
The choice of $\gamma$ and the frequency with which the thermostat
is called to work determine the relaxation rate, with which
the system adapts $T^{*}$. The version described in\cite{Allen87}
shows deviations of the achieved temperature for small numbers of
particles per cell, whereas our implementation exactly reproduces
$T^{*}$. The thermostat can even be extended to
particles of different mass i.e. colloid \emph{and} fluid particles
where the mass is used as weight factor for all velocities of the
simulation.
\end{subsection}

\begin{subsection}{Outlook: Shear}
There are several possibilities to shear the system. If one only has MD particles,
one can use moving walls either with a spring constant and a friction coefficient or with
direct hard reflections, where a moving wall is assumed and the reflection is calculated
in the moving frame fixed to the wall.
\\
These approaches of course neglect all effects (like pseudo wall slip), which appear close
to a wall in a shear experiment with a suspension. There, shear stress has to be applied to
the fluid which then drags the suspended particles. One way to implement this, is to add a small
velocity offset to all fluid particles which are reflected. Since this approach works well and
the colloidal particles are dragged by the fluid, we apply shear in this way to our system.
\end{subsection}
\end{section}

\begin{section}{Tests of the Simulation Code}
\begin{subsection}{Conservation of energy, velocity distributions}
\noindent
We have checked that the total energy is conserved in the molecular dynamics simulation if all
damping constants are switched off. Otherwise, or if the total energy even increases in spite of
damping constants, the MD time step has been chosen too large. In the SRD simulation energy is
conserved as well and if we use coupling method II also for the total system energy is
conserved within numerical accuracy. With coupling method I (where a thermostat is already
included in the coupling method) or if we switch on an additional thermostat energy will not exactly
be conserved but the system will reach a stable, i.e. equilibrated state. In that sense, total energy
(including thermal energy) will converge to a constant value.
\\
In SRD-Simulations without any embedded particles, the total energy contains only
the kinetic energy of the fluid particles. It is fully determined by the initialization
of the particle velocities. We can choose three uniformly distributed random numbers to initialize
the three velocity components for the fluid particles. In thermal equilibrium the distribution
should be a Gaussian, which in fact can be observed in our simulations after some tens of
SRD time steps.
If colloidal particles are included into the system, they should reach a thermal equilibrium,
at least as long as no external forces are applied. Damping terms would reduce fluctuations,
so, to check, if the colloidal particles reach the same temperature as the fluid particles, damping constants
have to be set to zero. Both distributions are shown in \fig\ref{fig_Velocity}.
They are both Gaussian with the correct temperature, even though for initialization
uniformly distributed random numbers (square well) had been used.
The tests are performed with both coupling methods. We have carried out simulations with particle
radii of $0.4\,\mu$m and $0.25\,\mu$m, where the Peclet number (for the simulations where
gravity is applied) is 0.11 and of course, it takes much longer to observe sedimentation.
\begin{figure}
a)\epsfig{file=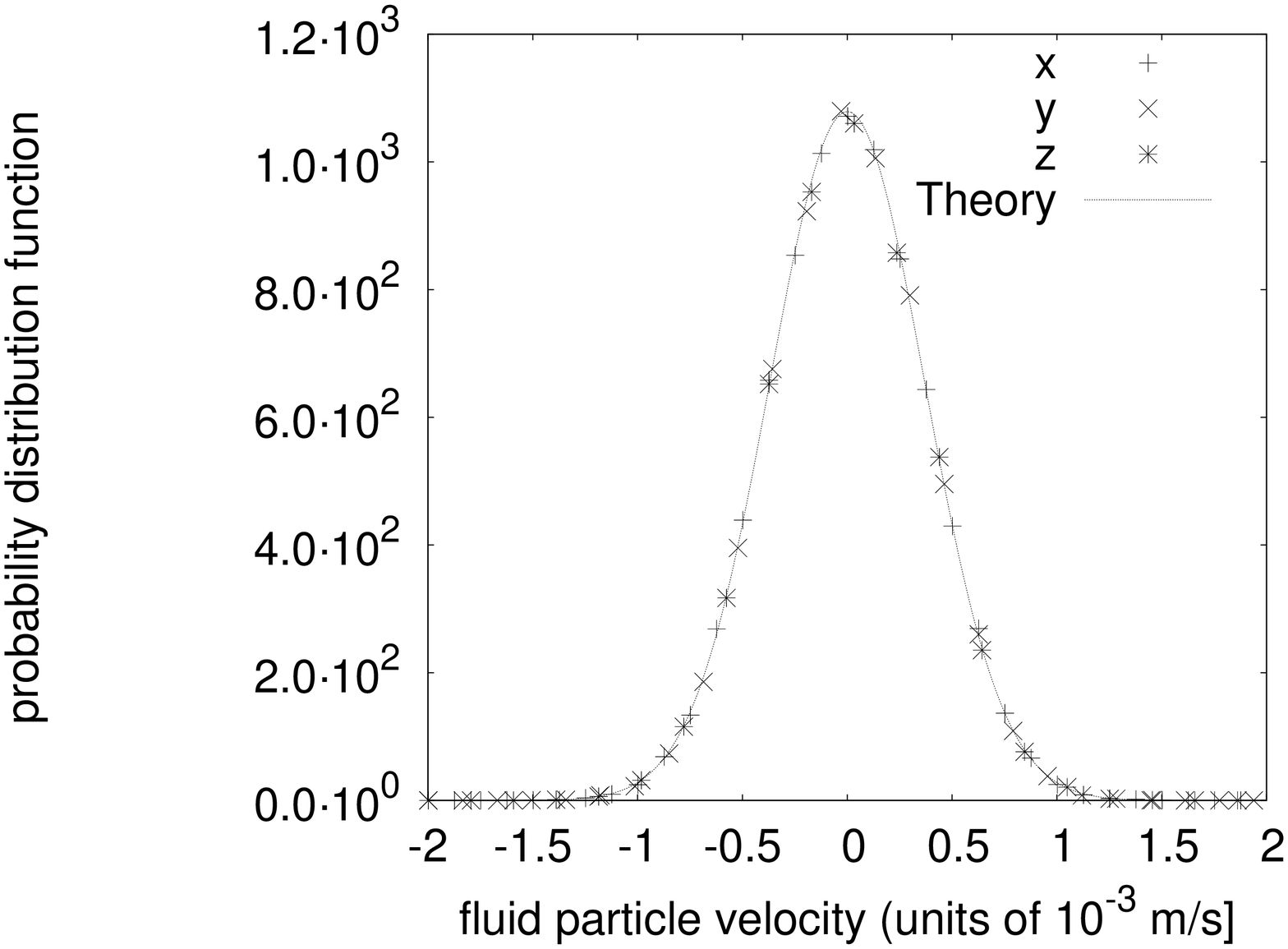,height=5cm} 
b)\epsfig{file=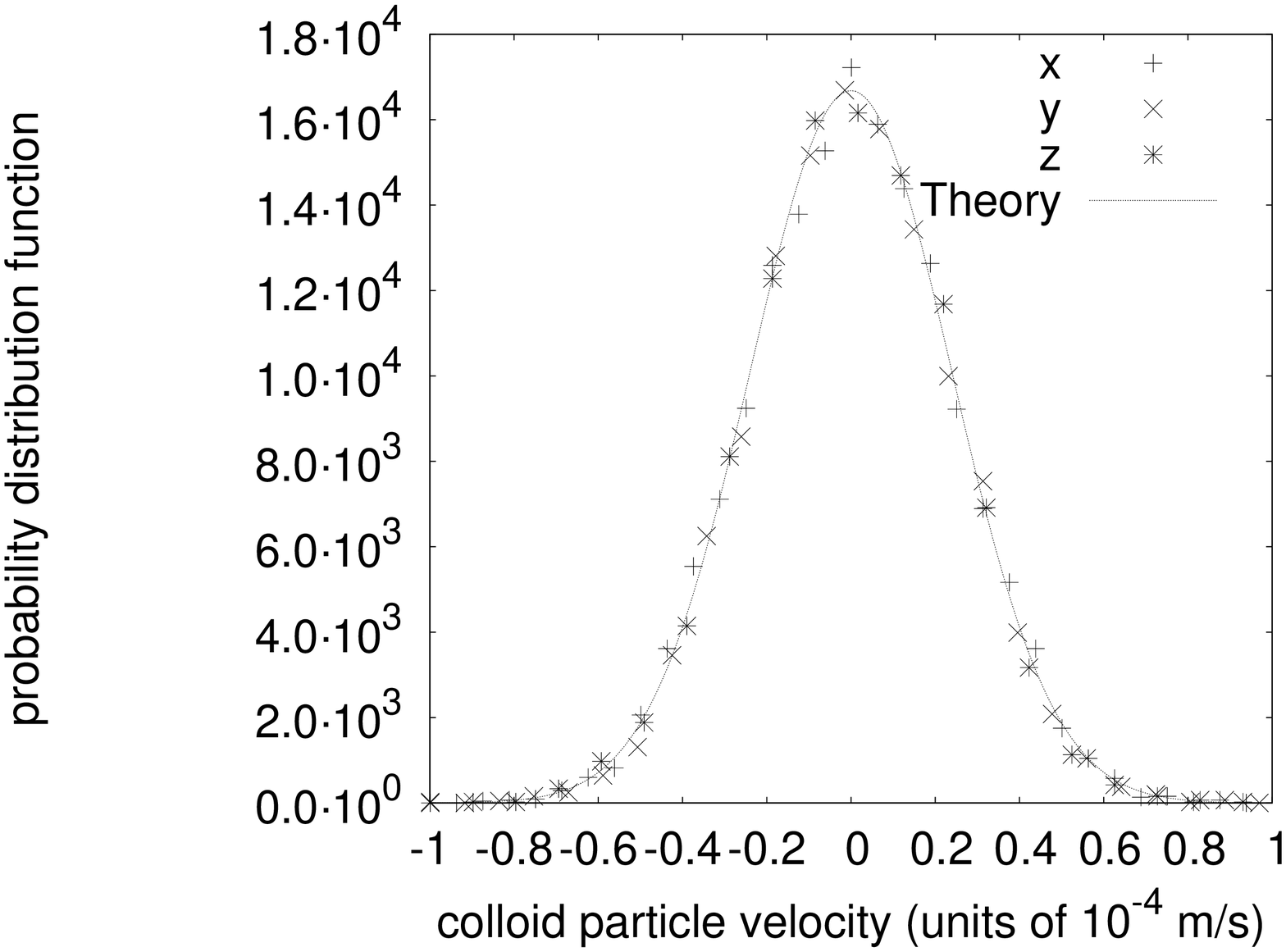,height=5cm}
\caption{Velocity distribution of fluid (a) and colloid (b) particles in a SRD simulation after thermalization.
Particle density is 3900 kg/m$^3$, the time step is 2.0$\,\mu$s, model temperature $T = 10.57$mK,
fluid particle mass $m_f = 1.0667 \cdot 10^{-18}$, particle diameter $d = 0.5\,\mu$m.
The theoretical Gaussian curve is plotted as well as the measured velocity distributions.
}
\label{fig_Velocity}
\end{figure}
\end{subsection}

\begin{subsection}{Viscosity}
The diffusion coefficient of suspended colloidal particles can be used to check if the
desired viscosity could really be achieved in the simulation. Using \eq(\ref{eq_Einstein})
we can, once we have measured $D$, calculate the kinematic viscosity $\nu$ and compare it
to the value we have used to determine the simulation parameters like the SRD time step.
We achieve a deviation of less than $20\%$ in a diluted system compared to the
theoretical value for an \emph{infinitely} diluted system.
Note that $D$ is a fixed number only in the limit of an infinitely diluted
system and only if the interaction potentials between the colloidal particles are exclusively
repulsive.

We are using two different methods, either the Green-Kubo-method or direct evaluation of
the mean square displacement. The first is even very accurate, if only few particles
are used, but consumes much computer time and memory because all particle velocities
have to be stored for all time steps used in the calculation. That means, for higher
volume fractions, it is more efficient just to sum up all the mean square displacements
within a given period of time.
To calculate $D$ using the Green-Kubo method one uses the following relation:
\begin{eqnarray}
  \label{eq_gxj}
    g_x(j) &=& \lim\limits_{I\rightarrow\infty} \frac{1}{I M_{\mathrm{Tot}}} \\ \quad \nonumber &&
    \sum\limits_{i=1}^{I}
    \sum\limits_{n=1}^{M_{\mathrm{Tot}}}v_{x,n}((i+j)\delta t) v_{x,n}(i\delta t) \\
  \label{eq_DxGK}
  D_{x} &=& \delta t \left(\frac{1}{2}g_x(0)+\sum\limits_{j=1}^\infty g_x(j)\right) 
\end{eqnarray}
where $M_{\mathrm{Tot}}$ is the total number of particles in the system, $I$ is the number of time
steps used to calculate the contribution $g(j)$. $v_{x,n}(i\delta t)$ denotes the $x$ component of
the velocity of particle $n$ in the $i$-th time step.
The sum in the expression for $D_x$ is in principle an infinite one,
but since the  contributions $g(j)$ decay with $j^{-3/2}$, one can truncate this sum after
some tens of terms. $D_y$ and $D_z$ can be calculated accordingly.
In \fig\ref{fig_Greenkubo}\,a we show the diffusion coefficient in each direction.
In numerical calculations it is impossible to evaluate an infinite sum.
In \eq(\ref{eq_gxj}) $I$ is limited at least by the total number of time steps
within the simulation and in \eq(\ref{eq_DxGK}) the sum therefore is not infinite either.
Since the contributions $g_x(j)$ become more and more inaccurate for larger $j$
we truncate the sum after $n$ terms and find that in our simulations for $n \approx 50$
the diffusion coefficient does not change anymore in a systematic way if $n$ is increased further.
In \fig\ref{fig_Greenkubo}\,b the last term of the sum is shown. For larger values, they fluctuate
due  to the finite sum in \eq(\ref{eq_gxj}) which leads to the inaccuracy in the right
part of \fig\ref{fig_Greenkubo}\,a. These fluctuations become smaller for longer simulation runs,
but do not change the value of the diffusion coefficient taken as an average from the center
part of \fig\ref{fig_Greenkubo}\,a\,\footnote{This can be seen as a smooth cutoff of the sum in
\eq(\ref{eq_DxGK}).}.
\\
For the mean square displacement in one direction during a time interval $\Delta t$ we calculate
\begin{equation}
  D_x = \frac{1}{2\Delta t M_{\mathrm{Tot}}}\sum\limits_{i=1}^{M_{\mathrm{Tot}}}\left(x_i(t+\Delta t)-x_i(t)\right)^2
\end{equation}
and $D_y$ and $D_z$ accordingly. For medium densities we have compared both methods and achieved the
same results within error bars. Depending on the number of particles, we use one of both methods.
\begin{figure*}
a)\epsfig{file=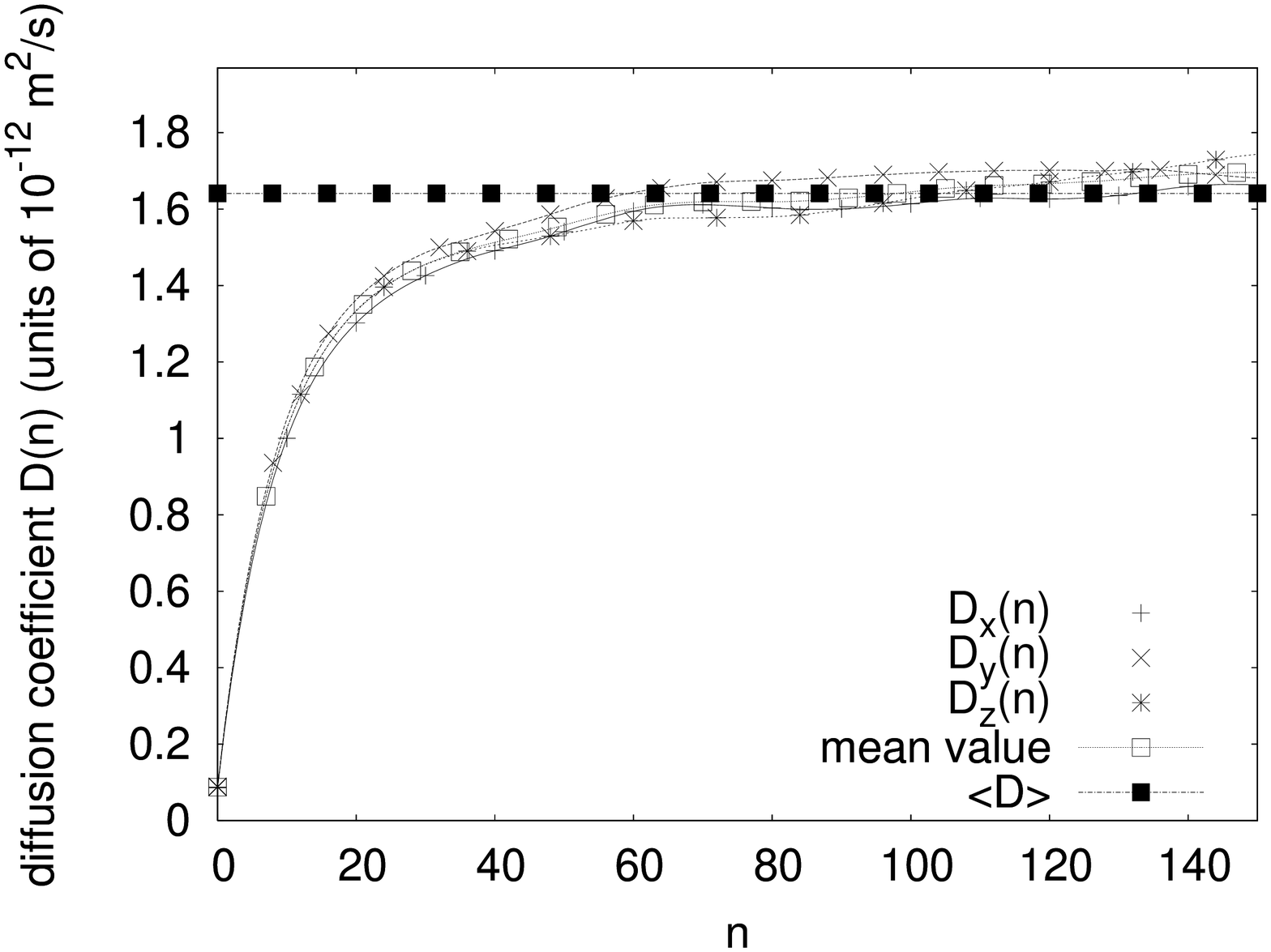,height=5cm} 
b)\epsfig{file=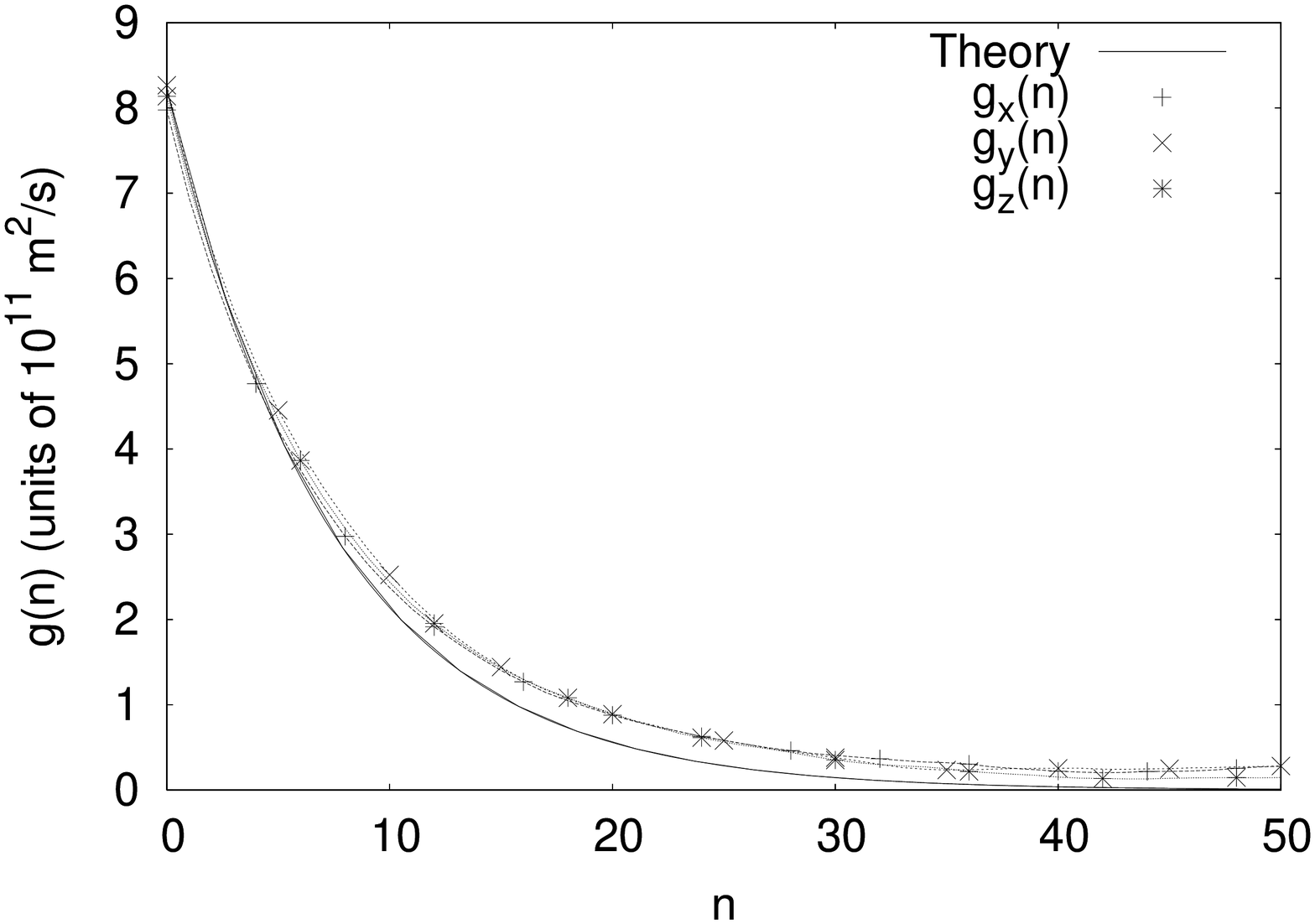,height=5cm}
\caption[a) Evaluation of $D$ using the Green-Kubo method: the plot shows the sum of
$\sum_{j=1..n} g_{x,y,z}(j)$ and the estimated $D$ \newline
b) the decay of the contributions $g_{x,y,z}(j)$. We have measured the
diffusion constant of soft spheres coupled with coupling II to the SRD at low volume
fractions.]
{
\parbox[t]{0.9\textwidth}{
a) Evaluation of $D$ using the Green-Kubo method: the plot shows the sum of \newline
$\sum_{j=1..n} g_{x,y,z}(j)$ and the estimated $D$  \newline
b) the decay of the contributions $g_{x,y,z}(j)$. We have measured the
diffusion constant of soft spheres coupled with coupling II to the SRD at low volume
fractions.
}
}
\label{fig_Greenkubo}
\end{figure*}

According to Richardson and Zaki\cite{RichardsonZaki}, the mean sedimentation velocity of particles
suspended in a liquid depends on the volume fraction $\phi$ as:
\begin{equation}
  \label{eq_ricZak}
   v_s (\phi) = v_{\infty} (1-\phi)^{l},
\end{equation}
with a typical exponent $l$ between $\approx 2.5$ and $4$ depending on the boundary conditions.
For periodic boundary conditions, Peclet number of $Pe = 1$
and Reynolds number $Re \ll 1$ we find an exponent of 3.5
(\fig\ref{fig_RichadsonZaki}) even when we use coupling method II,
where only long range hydrodynamic interaction can be calculated correctly.
A similar value is found for
$Pe = 2$ and $Pe = \frac{1}{2}$. Padding and Louis have found that the
exponent $l$ depends very weakly on the Peclet number\cite{Padding04}.
We have used here coupling method II, but some investigations have also been 
carried out using coupling method I. On the first view there is no big difference
apparent between the two coupling methods, at least, as long as, like in this test 
of our simulation code, no attractive forces are included. Our first results where
we have studied the peloid system in more detail are presented in the following section.
\begin{figure}
\epsfig{file=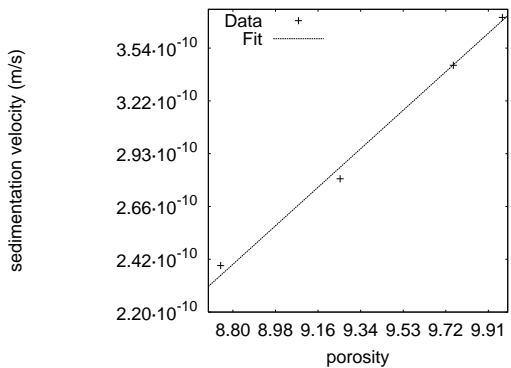,width=\linewidth}
\caption{Mean sedimentation velocity over porosity $(1-\phi)$ according to \eq(\ref{eq_ricZak}):
Measured values and fit curve in a log-log-plot. The Peclet number of this simulation is 1. Coupling
method II has been used for this plot.}
\label{fig_RichadsonZaki}
\end{figure}
\end{subsection}
\end{section}

\begin{section}{Results}
%
\begin{subsection}{Spacial correlation functions}
\begin{figure*}
\mbox{\epsfig{file=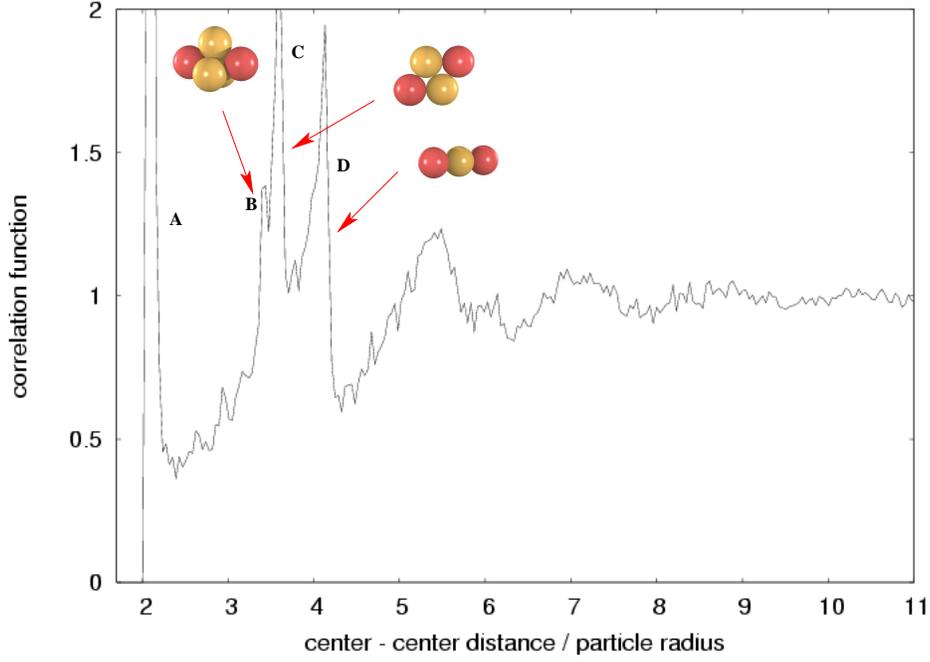,height=9cm}}
\caption{Correlation function of Al$_2$O$_3$ for $\Psi_0=50\,\mathrm{mV}$ and $\kappa=3\cdot10^{8}/\mathrm{m}$.
The potential is attractive, thus peaks (labeled by letters) can be identified and assigned to special local
configurations (see text).}
\label{fig_Correl}
\end{figure*}
\begin{figure*}
\epsfig{file=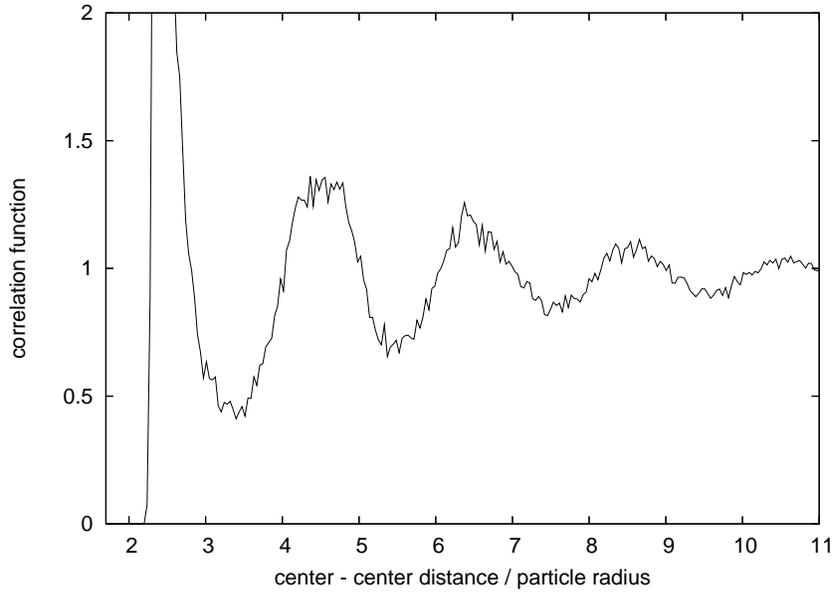,height=8cm}
\caption{Correlation function of Al$_2$O$_3$ for $\Psi_0=50\,\mathrm{mV}$ and $\kappa=7.3\cdot10^{7}/\mathrm{m}$.
repulsive potentials. One can see oscillations caused by the excluded volume.}
\label{fig_Correl7_3e7}
\end{figure*}
\noindent
For our production runs, we have simulated volume fractions of 7, 14, 21, 28 and 35\%
in a cube with an extension of ($6\,\mu$m)$^3$. Therefore are 231, 462, 693, 924 and
1155 colloidal particles respectively and $2.0 \cdot 10^5$ fluid particles necessary.
\\ 
We have evaluated the particle-particle correlation function. For attractive potentials
several sharp peaks can be observed and we assign them to distinct local orders of
particles. Oscillations can be found in the correlation function. They are caused by
exclusion of volume. In the case of attractive forces they are less pronounced than
if mainly repulsive interaction is present, see \fig\ref{fig_Correl} and
\fig\ref{fig_Correl7_3e7}.
\\
In \fig\ref{fig_Correl} the particles cluster due to their attractive potentials and
form stable configurations. The diameter of the particles is $5\cdot 10^{-7}\,$m.
There is a sharp peak in the spacial correlation function of the particle centers at exactly
that distance $2R$, where two particles touch each other in the very left part of the plot (A). Then, for
larger particle separation, the correlation function starts to grow and drops suddenly after a peak at
$1\,\mu\mathrm{m}$ (D), which is twice the diameter ($4R$). This is the contribution of two particles
touching the same third particle. The distance between them depends on the angle, which
they form with the particle in the middle, but, it is at last twice the diameter,
when they are in a straight line, which explains the sudden drop of the correlation function.
If several particles stick together, the straight line is stabilized. This explains the
peak at the end of this section of the correlation function.
\\
Two more peaks can clearly be assigned
to configurations: One of them is from two particles touching two other particles, which themselves
touch each other (C). There again the case of all particles being in the same plane can be stabilized
by other particles surrounding them. The particles under consideration are then separated by a distance
of $2R\sqrt{3}$. But of course, bending this configuration is still a degree
of freedom which brings the two particles slightly closer to each other. Thus their contribution
to the correlation function is shifted downward.
\\
The fourth peak at $\frac{4}{3}R\sqrt{6}$ reflects two particles, both touching three particles,
which themselves are touching each other and define a plane (B). There is no freedom
anymore for the two particles touching all the three of them at the same time. One can place one
of them at one side of the plane and the other one at the other side.
\\
When the potentials are mainly repulsive and the minimum caused by the van der Waals attraction
is only a fraction of $k_{\mathrm{B}}T$, the spatial correlation function looks completely
different, as depicted in \fig\ref{fig_Correl7_3e7}: The peaks described in the previous paragraphs
have disappeared here. The primary peak has moved to a slightly larger distance, since the
repulsive potential hinders the particles from touching each other.
\\
In \fig\ref{fig_PotCorr} we compare the correlation function of \fig\ref{fig_Correl7_3e7}
with the potential used for that simulation. The maximum of the correlation function coincides
with the minimum of the potential, but, as the minimum is not very sharp, the particles are not
restricted to fixed geometries and are in a steady process of rearrangement which results in
broader peaks. This process could also be studied by evaluating the velocity correlation function
for the colloidal particles which is related to the viscosity of the sample.
The correlation of particles which are several diameters apart is still remarkable,
as it is transmitted by the particles in between.
The oscillations of the correlation function can be understood as a formation of layers where the
probability of finding a particle in a certain layer is higher than in between.
\begin{figure}
\mbox{\epsfig{file=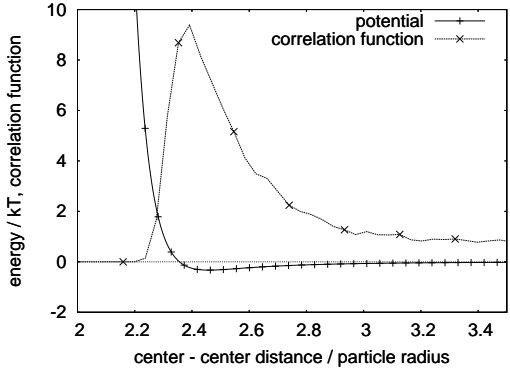,width=\linewidth}}
\caption{Plot of the correlation function of \fig\ref{fig_Correl7_3e7} together with the potential
used in this simulation. One can see that the maximum of the correlation function occurs for the
distance, at which the very shallow secondary minimum of the potential is located.
}
\label{fig_PotCorr}
\end{figure}
\end{subsection}
%
\begin{subsection}{Shear}
We have carried out simulations with shear and gravity. For the particles the boundaries in
$z$ direction were closed, gravity was applied in negative $z$-direction only to the
colloidal particles.
For the fluid particles the boundary in $z$-direction was closed as well and additionally
a velocity offset was added to apply a shear in $x$-direction. Boundaries for fluid and for
particles were periodic in $x$- and $y$-direction. Velocity distribution functions
have been evaluated. For the cases we investigated, after a transient they are all Gaussian
(\fig\ref{fig_ShearedColloids}).
\begin{figure}
\epsfig{file=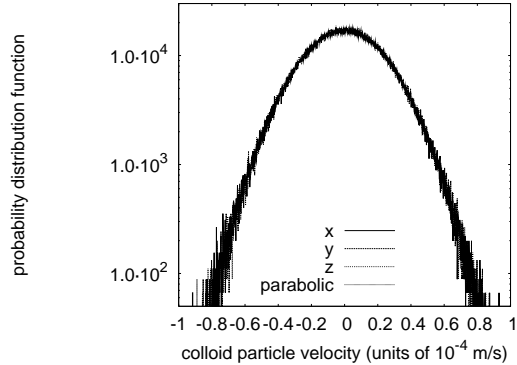,width=\linewidth} 
\caption{Velocity distribution of colloidal particles for each direction. Semi-log-plot where deviations from a
Gaussian would be visible by deviations from a parabolic profile.}
\label{fig_ShearedColloids}
\end{figure}

%
\begin{figure*}
\mbox{\epsfig{file=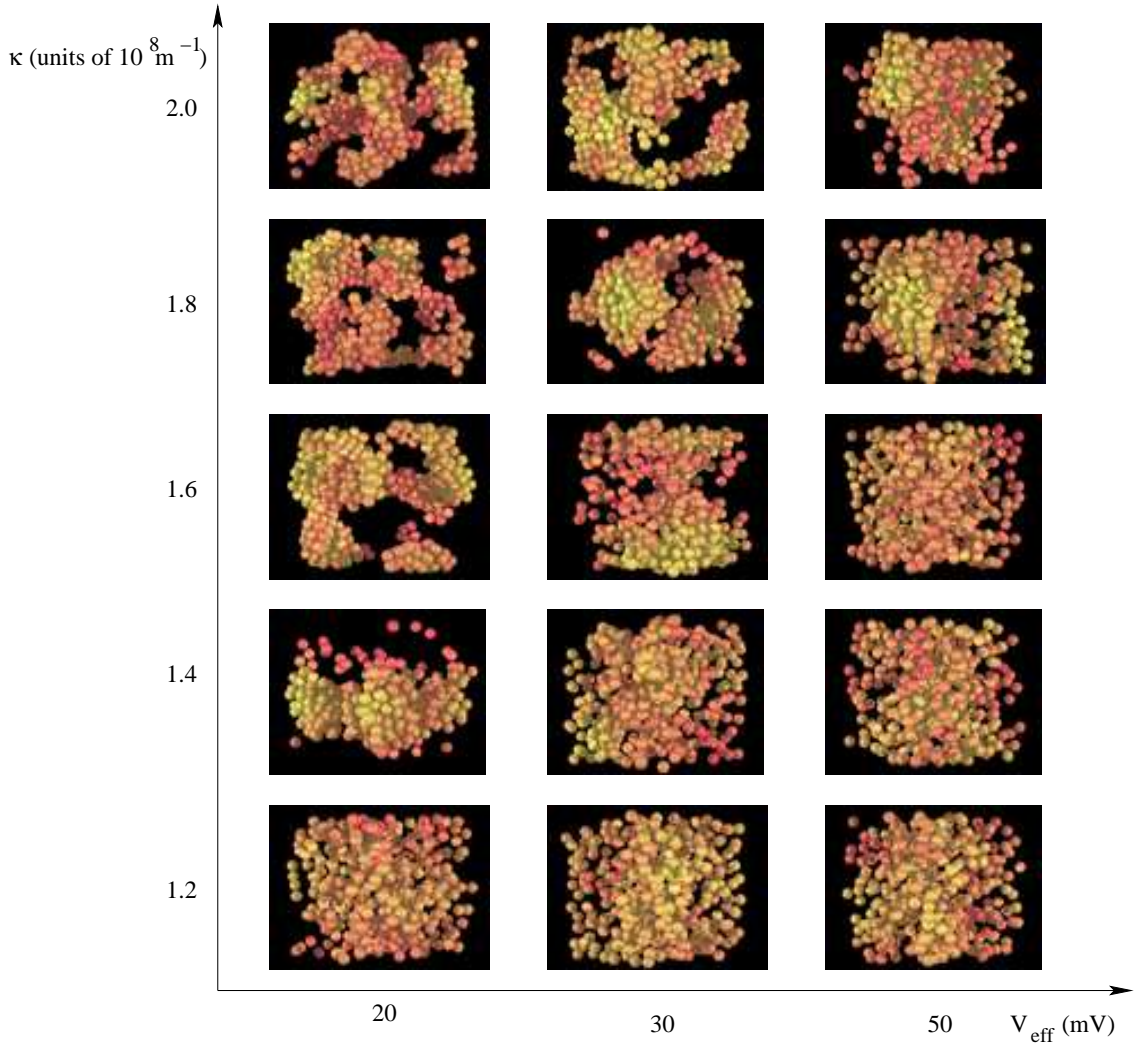,width=\textwidth}}
\caption{ Snapshots from the phase diagram of Al$_2$O$_3$: For the DLVO-Potentials with different
effective surface charge and different screening length one can either observe cluster formation or single
particles in suspension. The simulation was done at room temperature for 1 second of real time and a particle
diameter of $0.5\,\mu$m. Gravity has not been applied here. The pictures are corresponding to the
values written on the axis. For this figure we have chosen the simulation runs for 14\% volume fraction
with 462 colloidal particles.}
\label{fig_Phasediag}
\end{figure*}
\end{subsection}
%
\begin{subsection}{Phase diagram}
We have explored the phase diagram for Al$_2$O$_3$ with respect to screening length and
effective surface potential. We could identify the regions of suspended single particles and
of flocculation (\fig\ref{fig_Phasediag}). The transition between these two regions
depends on both parameters, Debye screening length \emph{and} effective surface potential.
It is known that the pH-value determines the effective surface potential $\Psi_0$,
and that salt concentration and pH-value determine the Debye screening length $\kappa$\,\cite{Wang99}.
Exact relations between salt concentration and pH-value on one side and $\kappa$ and
$\Psi_0$ on the other side are not known a priori for the parameter ranges of our
suspensions. There are approximations for very diluted systems and low salt concentrations.
It is known that for Al$_2$O$_3$ the surface potential becomes zero for $pH\approx 8.7$\cite{Huetter99}.
However, a phase transition between clustering in the upper left part of \fig\ref{fig_Phasediag}
and a suspended regime in the lower right part can be found in the simulations in analogy to the
experiment. The spatial correlation function can be evaluated for all the simulated cases and it
can be used as a tool to identify the two regions of the phase diagram.
\\
\figs\ref{fig_CorrelPlot1}--\ref{fig_CorrelPlot4} show selected examples of correlation functions
for different parameter sets. The first and second graph refer to a volume fraction $\Phi=14\%$
which also has been used for the phase diagram of \fig\ref{fig_Phasediag}. In \fig\ref{fig_CorrelPlot1}
the correlation function has been plotted for every other image of the left column in the phase diagram 
in \fig\ref{fig_Phasediag}. One can see that for suspended particles only the first peak can be found
in the correlation function. The secondary minimum in the potential causes the particles to
glue for short times before they continue with their diffusion process. With increasing $\kappa$
the secondary minimum approaches the particle surface, and therefore the main peak is shifted
to smaller distances. At the same time it becomes deeper so that clusters are formed and more peaks
occur. The peak at a distance of $2R\sqrt{2}\approx 3R$ disappears again, when the attraction becomes
stronger since this is a meta stable configuration of particles forming an octahedron.
\fig\ref{fig_CorrelPlot2} corresponds to the first row of images of \fig\ref{fig_Phasediag}.
In this case the depth of the secondary minimum is adjusted by changing the effective surface potential.
Again the transition between clustering regime and suspension can be observed. The potentials
used here are among the ones plotted above in \fig\ref{fig_Potentials}\footnote{In current simulations
we did not yet distinguish between clustering in the primary or secondary minimum.}.
In \fig\ref{fig_CorrelPlot3} and \ref{fig_CorrelPlot4} the dependence of the correlation function
on the volume fraction can be seen. In both cases long range correlations become more pronounced
with increasing volume fraction. This is shown for the suspended regime (\fig\ref{fig_CorrelPlot3})
and for the clustering regime (\fig\ref{fig_CorrelPlot4}), where the transition between the two
cases presented here is achieved by a variation of $\kappa$ by only $10\%$.

\begin{figure}
\mbox{\epsfig{file=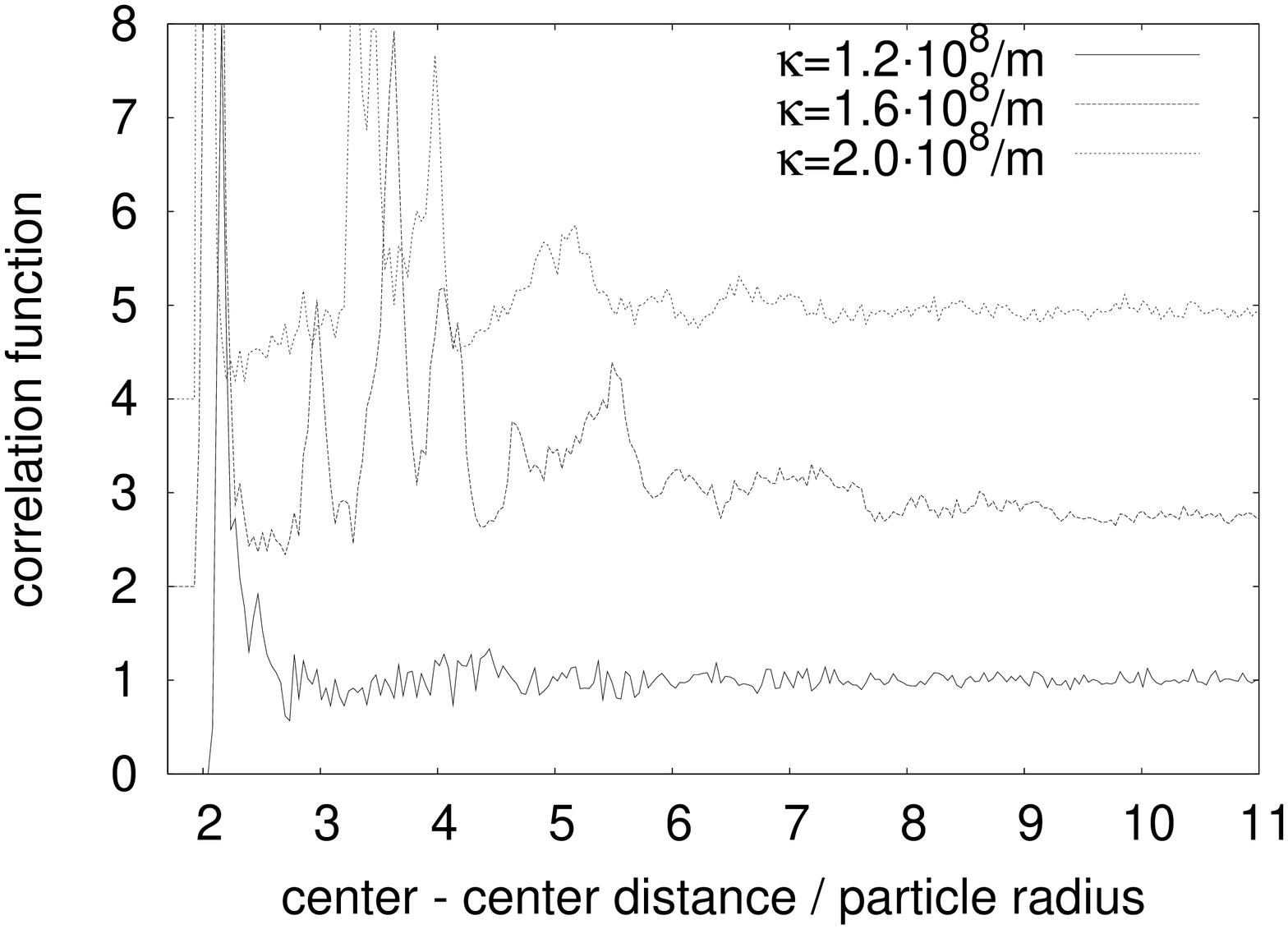,width=\linewidth}}
\caption{ Correlation function and its dependence on the inverse Debye screening length\,$\kappa$.
$\Psi_0=20\,\mathrm{mV}$ and $\Phi=0.14$ have been kept constant. For shorter Debye screening lengths
the attractive force becomes stronger and leads to clustering, which is reflected in the appearance of peaks.
The single curves have been shifted with respect to each other.}
\label{fig_CorrelPlot1}
\end{figure}

\begin{figure}
\mbox{\epsfig{file=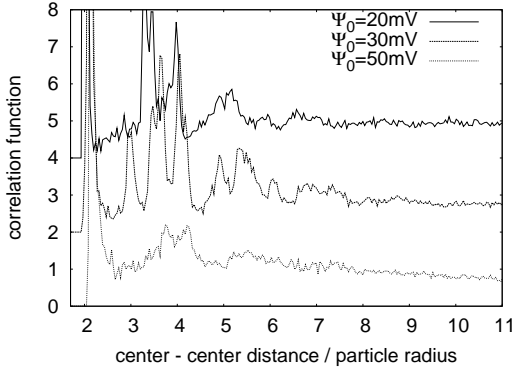,width=\linewidth}}
\caption{ Correlation function and its dependence on the effective surface potential $\Psi_0$.
$\kappa = 2 \cdot 10^{8}\,\mathrm{m}^{-1}$ and $\Phi=0.14$ have been kept constant.
The higher the effective surface potential, the stronger the attraction force and clustering
can be seen in the growing peaks.}
\label{fig_CorrelPlot2}
\end{figure}

\begin{figure}
\mbox{\epsfig{file=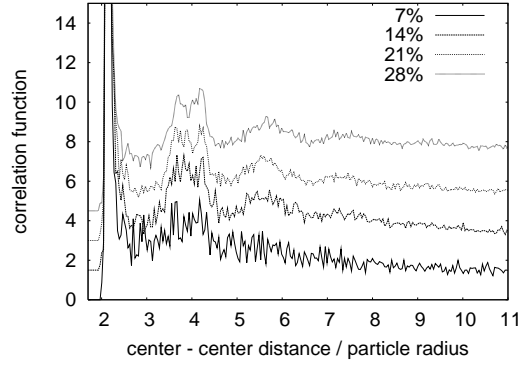,width=\linewidth}}
\caption{ Correlation function and its dependence on the volume fraction $\Phi$.
Effective surface potential $\Psi_0=20\,\mathrm{mV}$ and $\kappa = 1.4 \cdot 10^{8}\,\mathrm{m}^{-1}$
have been kept constant. For center-center distances between six and eight particle radii
broad peaks start to appear for larger volume fractions.}
\label{fig_CorrelPlot3}
\end{figure}

\begin{figure}
\mbox{\epsfig{file=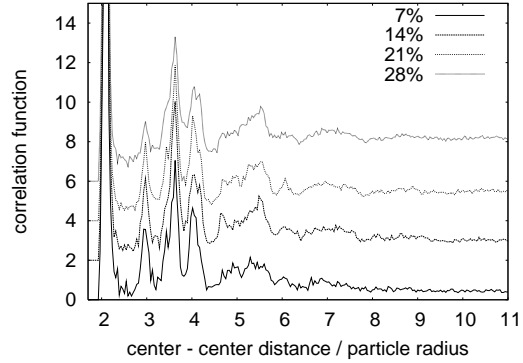,width=\linewidth}}
\caption{ Correlation function and its dependence on the volume fraction $\Phi$.
Effective surface potential $\Psi_0=20\,\mathrm{mV}$ and $\kappa = 1.6 \cdot 10^{8}\,\mathrm{m}^{-1}$
have been kept constant. Due to a small change in $\kappa$ with respect to \fig\ref{fig_CorrelPlot3}
one can cross the phase border between suspended particles and clustering regime. Also here long range
correlations become more pronounced for high volume fractions.}
\label{fig_CorrelPlot4}
\end{figure}
\end{subsection}
%
\begin{subsection}{Diffusion}
We measured the diffusion coefficient of colloidal particles with attractive potentials. In
\fig\ref{fig_Diffusion} we show the diffusion coefficient for Al$_2$O$_3$ with an effective
surface potential of $\Psi_0=50\,\mathrm{mV}$ and
an inverse Debye screening length of $\kappa = 2 \cdot 10^{8}\,\mathrm{m}^{-1}$ for room temperature.
One can see
that the mobility of the particles decays since a cluster formation process takes place and the
particles in the cluster are relatively fixed. The remaining mobility consists of two parts:
Particles can still, with a non vanishing probability, leave the cluster by thermal activation
and the cluster itself can take part in a diffusion process, it can vibrate or be deformed --
all of these are processes which are taking place on much longer time scales than the single
particle diffusion.
By studying the dependency of the diffusion coefficient on the potentials and on the
volume fraction, one might be able to find an answer to the question, which of these processes
is important for the dynamics of the system in which part of the phase diagram of \fig\ref{fig_Phasediag}.
\end{subsection}
\end{section}

\begin{section}{Conclusion}
\noindent
We have shown that by combining a Stochastic Rotation Dynamics and a Molecular Dynamics simulation
it is possible to study dense colloidal suspensions. We have explained how to determine effective
parameters for the simulation (box size $a$, simulation time step $\delta t$,
number of fluid particles per box $M$\dots). It is possible to relate the simulation
to very distinct experimental conditions since all parameters (density, temperature,
potentials\dots) which enter into the description are scaled in a well defined manner.
We have presented first results which demonstrate the power of the model. We have demonstrated that
the Richardson-Zaki law is reproduced already with the simple and fast coupling method II and
we have studied the dependence of the pair correlation function on the shape of the interaction
potentials. We have shown how one can distinguish if for given Debye screening length $\kappa$,
effective surface potential $\Psi_0$ and Hamaker constant $a_{\mathrm{H}}$ if the
system is in the clustering or suspended regime.
\\
We are planning to carry out detailed investigations of the properties described in the two preceding
sections (diffusion coefficient, correlation functions, sedimentation velocity) as well
as cluster size and shape. Then these quantities can be analyzed under shear, their
dependence on the shear rate, and the shear viscosity of the suspension, containing the
fluid and the particles, which both contribute to a complex shear viscosity.
\begin{figure}
\mbox{\epsfig{file=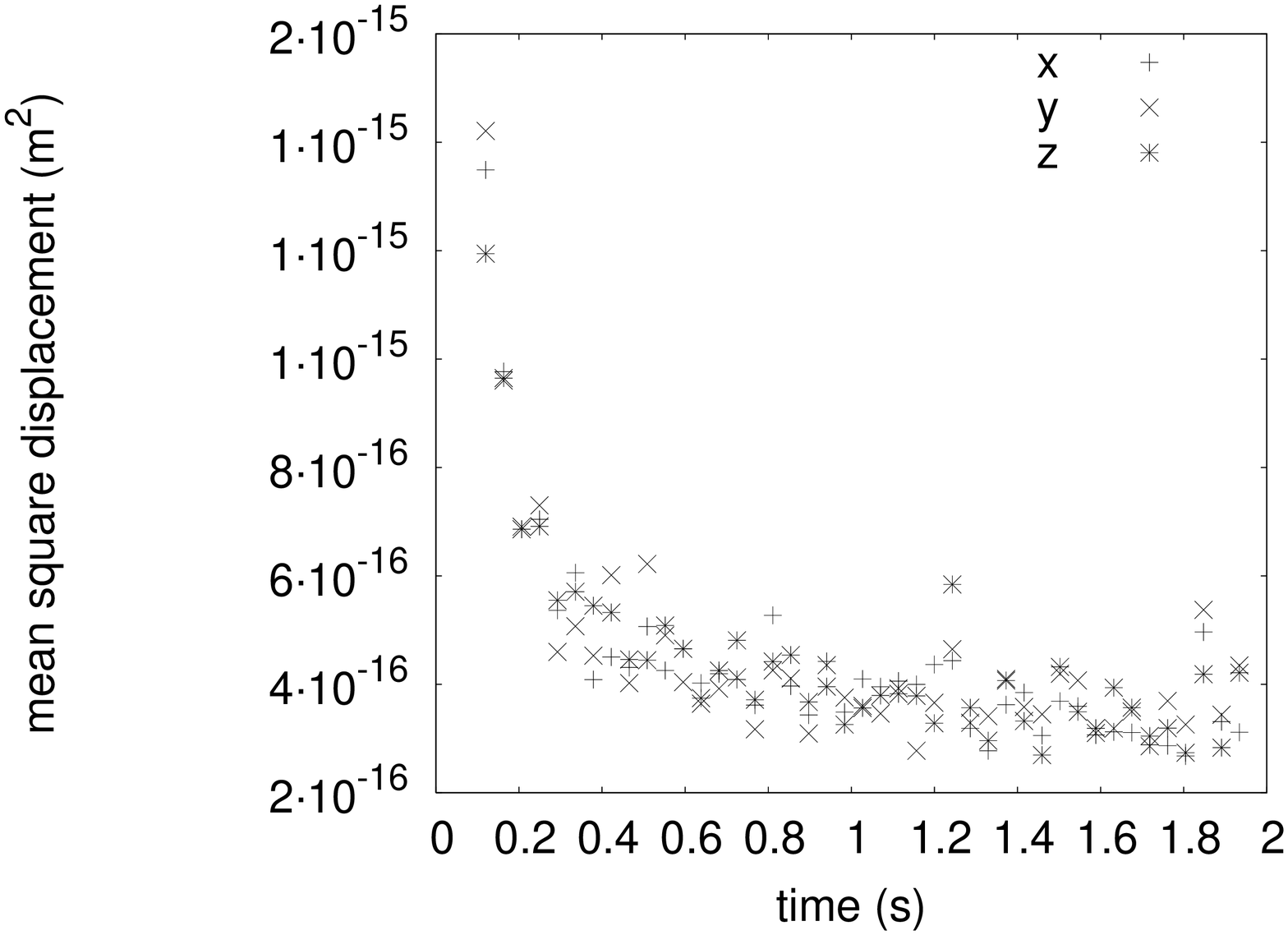,width=\linewidth}}
\caption{ Mean square displacement (projection on each of the axis). For this simulation $\Psi_0 = 50\,\mathrm{mV}$
and $\kappa = 3 \cdot 10^8\,/\mathrm{m}$ have been used. First particles move by diffusion, are attracted
and then form clusters of larger size and lower mobility, which can be observed in a decay of the diffusion
length for a given period of time. We have simulated 1155 particles in a cube with $6\,\mu$m extension, which
results in a volume fraction of $35\%$.}
\label{fig_Diffusion}
\end{figure}
\end{section}

\vspace{-5mm}
\begin{acknowledgments}
This work has been financed by the German Research Foundation (DFG) within the project
DFG-FOR 371 "Peloide". We thank G. Gudehus, G. Huber, M. K\"ulzer, L. Harnau, J. Reinshagen,
S. Richter, and M. Bier for valuable collaboration. We thank M. Strauss, A. Komnik, E. Tuzel,
D. M. Kroll, A. J. Wagner, Y. Inoue and M.~E. Cates for helpful discussions.
T. Ihle thanks the SFB 404, project A7, of the DFG and ND EPSCoR 
through NSF grant EPS-0132289 for financial support.
\end{acknowledgments}


\end{document}